\documentclass[superscriptaddress,amsmath,amssymb,aps,pra,twocolumn,floatfix]{revtex4-2}

\usepackage{graphicx}
\usepackage{bm}
\usepackage[colorlinks, linkcolor=mycol, citecolor=mycol, urlcolor=mycol, breaklinks]{hyperref}
\usepackage{braket}
\usepackage{bbold}
\usepackage{xcolor} 
\usepackage{comment}
\usepackage{physics}
\usepackage{verbatim}

\newcommand{\mi}{ {\rm i} }
\newcommand{\me}{ {\rm e} }
\newcommand{\id}{\mathbb{1}}

\usepackage{amsmath}
\usepackage{amssymb}
\definecolor{mycol}{RGB}{10,55,130}
\definecolor{mycol2}{RGB}{50,187,108}

\begin{document}
\title{Comparing bipartite entropy growth in open-system matrix-product simulation methods} 

\author{Guillermo Preisser}
\affiliation{CESQ and ISIS (UMR 7006), CNRS and Universit\'{e} de Strasbourg, 67000 Strasbourg, France}

\author{David Wellnitz}
\affiliation{JILA, National Institute of Standards and Technology and Department of Physics, University of Colorado, Boulder, Colorado, 80309, USA}

\author{Thomas Botzung}
\affiliation{Université Grenoble Alpes, CNRS, LPMMC, 38000 Grenoble, France}

\author{Johannes Schachenmayer}
\thanks{schachenmayer@unistra.fr}
\affiliation{CESQ and ISIS (UMR 7006), CNRS and Universit\'{e} de Strasbourg, 67000 Strasbourg, France}

\date{\today}

\begin{abstract}
The dynamics of one-dimensional quantum many-body systems is often numerically simulated with matrix-product states (MPS). The computational complexity of MPS methods is known to be related to the growth of entropies of reduced density matrices for bipartitions of the chain. While for closed systems the entropy relevant for the complexity is uniquely defined by the entanglement entropy, for open systems it depends on the choice of the representation. Here, we systematically compare the growth of different entropies relevant to the complexity of matrix-product representations in open-system simulations. We simulate an XXZ spin-1/2 chain in the presence of spontaneous emission and absorption, and dephasing. We compare simulations using a representation of the full density matrix as a matrix-product density operator (MPDO) with a quantum trajectory unraveling, where each trajectory is itself represented by an MPS (QT+MPS). We show that the bipartite entropy in the MPDO description (operator entanglement, OE) generally scales more favorable with time than the entropy in QT+MPS (trajectory entanglement, TE): i) For spontaneous emission and absorption the OE vanishes while the TE grows and reaches a constant value for large dissipative rates and sufficiently long times; ii) for dephasing the OE exhibits only logarithmic growth while the TE grows polynomially. Although QT+MPS requires a smaller local state space, the more favorable entropy growth can thus make MPDO simulations fundamentally more efficient than QT+MPS. Furthermore, MPDO simulations allow for easier exploitation of higher-order Trotter decompositions and translational invariance, allowing for larger time steps and system sizes.
\end{abstract}

\maketitle

\section{Introduction}

In recent years, experimental developments have made it possible to analyze and control almost fully coherent quantum many-body dynamics~\cite{cirac2012goals}, in particular for effective spin-1/2 models, e.g.~with trapped ultracold atoms or molecules in optical lattices~\cite{bloch2012quant,gadway2016strong}, Rydberg excitations~\cite{low2010experimental,adams2019rydberg,browaeys2020many,morgado2021quantum}, or ion traps~\cite{blatt2012quant}. The possibility of controlled quantum simulations on such platforms promises various practical applications that may lead to a quantum advantage compared to classical simulations~\cite{daley2022practical}.

\smallskip

The study of entanglement in many-body models has been a long-standing theory quest~\cite{amico2008entanglement,eisert2010colloquium}, especially  its growth dynamics (e.g.~\cite{calabrese2005evolution,fagotti2008evolution,alba2018entanglement}). The study of entanglement is of practical interest in one dimension, where the growth of the bipartite entanglement entropy $S$ (see Sec.~\ref{sec:model_mpentr} for definitions) is directly connected to the question of whether dynamics can be efficiently simulated on a classical computer. This connection is made via the concept of matrix-product states (MPSs) \cite{vidal2004efficient,schuch2008entropy, verstraete2008matrix,schollwock2011density-matrix,paeckel2019time-evolution}. An MPS is a numerical decomposition of a many-body state vector into a product of $\chi \times \chi$ matrices, where the entries of the matrices are local kets. For such a representation, the bipartite entanglement entropy is limited to $\max[S] = \log_2(\chi)$. Consequently, to represent a physical state $\ket{\psi (t)}$ with entanglement entropy $S(t)$ as MPS over time, the matrix size (or ``bond dimension'') has to grow at least as $\chi \propto 2^{S(t)}$. An evolution where $S$ increases faster than logarithmically can therefore be considered computationally inefficient.

\smallskip

Every experiment features at least small couplings to the environment, and should therefore be treated as an open quantum system described by a density matrix $\hat \rho$. Then, the definition of entanglement needs to be adapted: A bipartition of the open system is effectively a tripartition of a larger (closed) system, with the environment acting as third party, and bipartite entropies do not necessarily indicate entanglement. However, analogously to MPSs for pure states, also a matrix-product decomposition of the density matrix can be defined, known as a density matrix-product density operator (MPDO)~\cite{zwolak2004mixed,orus2008infinite,werner2016positive,GuthJarkovsky_Effici_2020,weimer2021simulation}. Equivalently to the pure state case, the so-called operator space entanglement entropy, or simply ``operator entanglement'' (OE) \cite{zanardi2000entangling,zanardi2001entanglement,wang2002quantum,prosen2007operator, dubail2017entanglement,zhou2017operator,jonay2018coarse, alba2019operator,wang2019barrier,styliaris2021information,noh2020efficientclassical,landa2022nonlocal}
is linked to the efficiency of this MPDO representation.

\smallskip

\begin{figure}[t]
	\centering
	\includegraphics[width=\columnwidth]{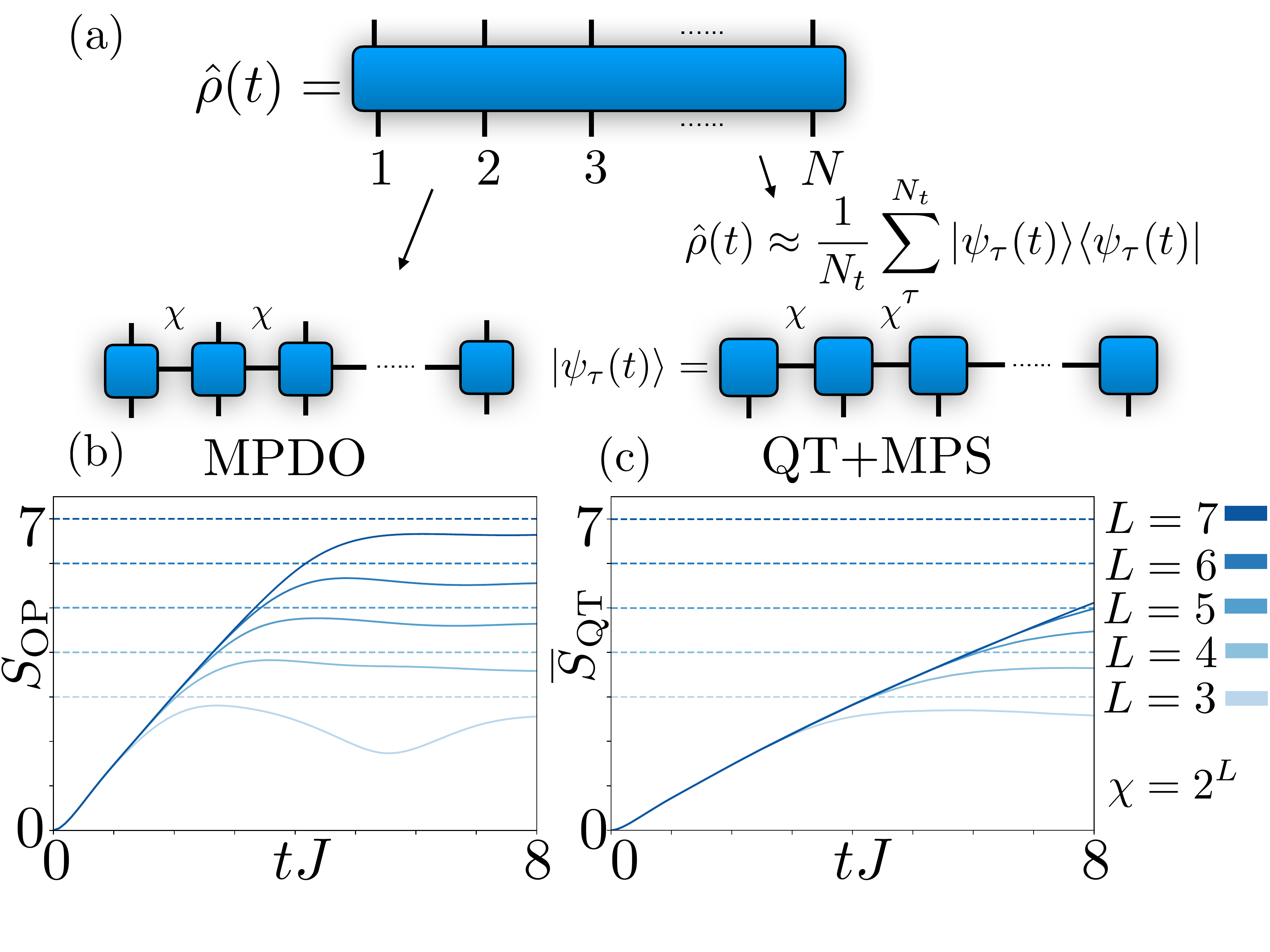}
	\caption{\textit{Overview }--- (a) Schematics of different numerical approaches for simulating master equation dynamics: i) decomposition of the density matrix into a matrix-product density operator (MPDO); ii) unravelling of the density matrix into matrix-product state (MPS) quantum trajectories. MPDO and MPS bond dimensions, $\chi$, limit operator entanglement (OE, $S_{\rm OP}$) and trajectory entanglement (TE, $\overline{S}_{\text{QT}}$), respectively (see text). (b/c) Evolution of OE and TE for a simulation of the master equation~\eqref{eq:master} with an initial N\'eel state. Results are shown in a small dissipation limit ($\gamma_\pm=0.01J$, $\gamma_z=0$), and entropies are computed for a splitting at the center (averaged over 11 center bonds for QT+MPS). Different lines correspond to simulations with increasingly large bond dimensions $\chi = 2^L$ and $L=3,\dots,7$, horizontal dashed lines indicate the entropy limits $\log_2(\chi)$. Statistical error bars are within the linewidths [$N=40$, for QT+MPS: $N_t = 400$].}
	\label{fig:intro}
\end{figure}

Alternatively, the density matrix can be nonuniquely decomposed as a statistical mixture of pure states [see Fig.~\ref{fig:intro}(a) for a sketch]. Each pure state can then be expressed as an MPS, and the efficiency of such a representation is determined by the bipartite entanglement entropies in the individual states. In order to compute time evolution, pure state trajectories can be stochastically evolved with a technique known as the quantum Monte Carlo wave-function method or quantum trajectories (QTs)~\cite{dalibard1992wave,dum1992monte,carmichael2009open}. This approach has been successfully applied in many simulations with MPS~\cite{daley2014quantum,weimer2021simulation}. We denote it as QT+MPS, and the average bipartite entropy of the trajectories as ``trajectory entanglement'' (TE). In general, due to the smaller local state space, a QT+MPS approach seems favorable compared to an MPDO, in particular for computations where only small numbers of trajectories are required for statistical convergence (e.g.~for computing evolution of simple local observables). However, a decreased entropy scaling could make the MPDO approach fundamentally more efficient since it can significantly relax the requirements on bond dimensions $\chi$~\cite{prosen2007efficiency}.

\smallskip

The answer to the question about the relative performance of the two different approaches depends on the model Hamiltonian, the initial state, the nature of dissipation, and the observable of interest. Which of the two approaches is preferable was analyzed for several specific setups in previous works, e.g., for a Bose-Hubbard model subjected to dephasing~\cite{bonnes2014superoperators}, for disordered fermionic hopping models with particle loss and dephasing~\cite{vannieuwenburg2017dynamics}, and for computations of two-time correlations in an XXZ model with spontaneous emission or dephasing~\cite{wolff2020numerical}. For the Bose-Hubbard model discussed in Ref. ~\cite{bonnes2014superoperators} calculations were limited to relatively small systems up to 20 sites, which led to a saturation of both OE and TE to relatively small constant values quickly, with the averaged TE staying below the OE. For such scenarios it was argued that the QT+MPS approach is favorable, except for limits of large dephasing rates. In the latter case QT suffers from extremely small time-step requirements. In contrast, for the disordered fermionic model in Ref.~\cite{vannieuwenburg2017dynamics}, the disorder leads to a strong suppression of both TE and OE. In this setup it was then argued that an MPDO approach is more efficient. In Ref.~\cite{wolff2020numerical} a thorough comparison led to the conclusion that for the computation of two-time correlations the magnitude of the dephasing rates is the crucial factor that determines which approach is favorable, related to the small time-step problem in the QT method. It was furthermore pointed out that the initial state plays a very important role, since already the compression of an initially entangled state into the MPDO form can be very costly, rendering the approach impractical.

In addition to the previous works, here we now perform a systematic comparison of OE and TE scaling behavior at long times in sufficiently large systems. We focus on the generic XXZ model with spontaneous emission and absorption, and dephasing for a simple initial product N\'eel state. Our simulations reveal that OE and TE exhibit a strikingly different scaling.  Recent work in Ref.~\cite{vovk2021entanglement} compared the TE scaling for different master equation unraveling strategies. It was found that when using a homodyne measurement unraveling~\cite{tian1992quantum}, the TE scaling can fundamentally change (from volume to area law), depending on homodyne measurement parameters. Consequently an entanglement optimized trajectory method can exploit this~\cite{vovk2021entanglement}. It was shown that entanglement can thereby be reduced compared to the commonly used number measurement method (exhibiting also area law behavior). It is now also important to compare TE scaling laws with those of OE, which is the main objective of this paper.

\smallskip

In this paper we consider a generic XXZ model with different types of noise, and compare MPDO simulations with QT+MPS (number measurement scheme). While for small noise rates and sufficiently short times TE seems to be generally smaller than OE [see Fig.~\ref{fig:intro}(b) and (c)], ultimately, we find that for larger rates and sufficiently long times, the MPDO approach has fundamental advantages compared to the QT+MPS ansatz in terms of the entropy scaling: The OE growth is significantly reduced (at times longer than the inverse rate of dissipation).  In the case of spontaneous emission and absorption the OE vanishes, while the TE exhibits a growth and approaches a constant for large dissipation rates and long times. This implies that here QT+MPS produces fundamentally inefficient representations with entangled trajectory states. We derive an analytical value for the constant TE in a large dissipation limit. We find a similar scenario for dephasing: Here, OE scales logarithmically with time (as long as magnetization conservation is not broken)~\cite{wellnitz2022rise,medvedyeva2016influence}, while TE increases as a power law. In addition to the fundamentally different entropy scaling, we also point out several practical differences in the implementation of both methods that make the MPDO approach more attractive. The latter allows us e.g.,~to implement higher-order Trotter decomposition and to exploit translational invariance easily. We also point out that for the correct choice of basis, MPDO elements can be made real.

\smallskip

This paper is organized as follows: In Sec.~\ref{sec:model} we introduce our XXZ and noise models. In Sec.~\ref{sec:model_mpentr} we review the bipartite entropies associated to MPDO and QT+MPS, respectively. We compare the two approaches for the different noise models in Sec.~\ref{sec:decay} for spontaneous emission an absorption and in Sec.~\ref{sec:dephasing} for dephasing, respectively. In Sec.~\ref{sec:methods} we provide technical details on our numerical methods and analytical calculations. Finally, we provide a conclusion and an outlook in Section~\ref{sec:concl}.

\section{Model}
\label{sec:model}

In this paper we consider a one-dimensional spin-1/2 XXZ chain with Hamiltonian ($\hbar \equiv 1$ throughout this paper) 
\begin{align} \label{eq:xxz}
\hat H_{
\rm XXZ} = \sum^{N-1}_{i=1} \frac{J}{4} \bigg[-(\hat \sigma^x_i \hat \sigma^x_{i+1} + \hat \sigma^y_{i} \hat \sigma^y_{i+1}) + \hat \sigma^z_i \hat \sigma^z_{i+1}\bigg].
\end{align}
Here, $N$ is the number of spins, $\hat \sigma^{x,y,z}_i$ denote standard Pauli matrices defined in a local basis $\ket{\downarrow,\uparrow}_i$, and $J$ is the nearest-neighbor spin-coupling strength. For finite system size calculations we consider open boundary conditions. Note that in the main text we consider a fixed anisotropy in the XXZ model. However, we point out that the OE and TE exhibit interesting opposite behavior as a function of the ZZ interaction, see Appendix~\ref{sec:zz_dependence}.

Starting in a pure N\'eel product state,
 \begin{align} \label{eq:neel}
\hat \rho (t=0) = \bigotimes_{i=1}^{N/2}\ket{\uparrow}_{2i-1}\ket{\downarrow}_{2i}\bra{\uparrow}_{2i-1}\bra{\downarrow}_{2i},
\end{align}
we will consider open-system dynamics under a Lindblad master equation, i.e.~for a chain being coupled to a Markovian environment~\cite{carmichael2009open}

\begin{align} \label{eq:master}
\frac{d}{dt} \hat \rho = -\mi [\hat H_{\rm XXZ}, \hat \rho] + \sum_\eta \mathcal{D}[\hat L_\eta]\hat \rho.
\end{align}
Here, our dissipative channels are given by super-operators
\begin{align}
\mathcal{D}[\hat L_\eta]\hat \rho =   \hat L_\eta \hat \rho \hat L_\eta^\dag - \frac{1}{2} \{\hat L_\eta^\dag \hat L_\eta, \hat \rho \},
\end{align}
which are defined by the Lindblad jump operators $\hat L_\eta$. The curly bracket denotes the anti-commutator. Alternatively the master equation can be written in the form
\begin{align} \label{eq:master_alt}
\frac{d}{dt} \hat \rho = -\mi \left(\hat H_{\rm eff} \hat \rho - \hat \rho \hat H_{\rm eff}^\dag \right) + \sum_\eta \hat L_\eta \hat \rho \hat L_\eta^\dag
\end{align}
with the effective Hamiltonian $\hat H_{\rm eff} = \hat H_{\rm XXZ} + \hat H_{\rm nh}$ with non-Hermitian part $\hat H_{\rm nh} = -\mi \sum_\eta \hat L^\dag_\eta \hat L_\eta /2$. In the following we will consider two noise models: 

\medskip

\paragraph{Spontaneous emission and absorption --} In the first scenario we consider local incoherent transitions between the two states $\ket{\uparrow}_i \leftrightarrow \ket{\downarrow}_i$, defined respectively by the two jump operators for spin $i$:

\begin{subequations}
\label{dis}
\begin{align}
\hat L^+_i &= \sqrt{{\gamma_+}} \hat \sigma^+_i \label{eq:Lp} \\
\hat L^-_i &= \sqrt{{\gamma_-}} \hat \sigma^-_i. \label{eq:Lm}
\end{align}
\end{subequations}
Spontaneous emission $\ket{\uparrow}_i \to \ket{\downarrow}_i$ occurs for example naturally for atomic two-level systems coupled to the electron-magnetic vacuum. For optical transition frequencies, spontaneous absorption $\ket{\downarrow}_i \to \ket{\uparrow}_i$ is typically negligible, but can be engineered e.g.~by using a laser drive to a highly excited state that quickly relaxes to $\ket{\uparrow}_i$~\cite{zhu2015synchronization}. Here, we will mostly focus on the scenario where $\gamma_+ = \gamma_-$. This scenario conserves the mean magnetization of the initial N\'eel state, i.e.~at all times $\Tr(\hat S^z \hat \rho) = 0$ with $\hat S^z = \sum_i \hat \sigma^z_i$, which implies that the state explores a large Hilbert space throughout the evolution. Note that, in contrast, for the maximally imbalanced cases $\gamma_+=0 < \gamma_-$ or $\gamma_-=0 < \gamma_+$, the system relaxes to trivial steady states $\bigotimes_i \ket{\downarrow}_i$ or $\bigotimes_i \ket{\uparrow}_i$, respectively.

\medskip

\paragraph{Dephasing --} As a second noise model we consider an interaction with the environment that leads to dephasing, a loss of a definite phase relation between the two spin-states. Such a mechanism occurs when the environment effectively measures the spin-states without changing them. For example, for a spin-model realized with hard-core bosons in an optical lattice, such a process can be the dominant dissipative mechanism, due to atoms spontaneously absorbing and re-emitting light from the lattice lasers~\cite{pichler2010nonequilibrium}. In XXZ models, dephasing noise has been shown to be important~\cite{wolff2019evolution}, with further connections to Tomonaga-Luttinger liquids~\cite{bernier2020melting, buchhold2015nonequilibrium}. In the master equation, we describe dephasing by the local jump operators
\begin{align}
\hat L^z_i &= \sqrt{{\gamma_z}} \hat \sigma^z_i. \label{eq:Lz}
\end{align}
Note that for dephasing, the magnetization is associated with a strong symmetry, i.e.~it commutes with the Hamiltonian and all Lindblad operators $[\hat H, \hat S^z] = [\hat L_\eta, \hat S^z] = 0$. As a consequence, the state remains restricted to a single symmetry sector, here given by magnetization $S^z = 0$. This can be exploited to make numerical (MPS) simulations more efficient, and can lead to fundamentally interesting many-body physics~\cite{macieszczak2019coherence,rossini2021coherent,cai2013algebraic,medvedyeva2016exact,foss2017solvable,znidaric2015relaxation}. A direct consequence of the magnetization conservation is a long-time logarithmic growth of OE~\cite{wellnitz2022rise}.

\section{Matrix-Product Decomposition and bipartite entropy}
\label{sec:model_mpentr}

In order to exactly represent the $2^N \times 2^N$-dimensional density matrix of our system, in principle $(4^N-1)$ real numbers are needed. Because of the exponential dependence on $N$, exact diagonalization approaches become thus intractable even for moderately large $N \gtrsim 15$. However, if the entanglement between different subsystems is not too large, the density matrix may be efficiently represented as a matrix-product decomposition. For pure states, this decomposition is well known as an MPS. The MPS concept can be generalized to mixed states in several ways, and the appropriate measure of entanglement depends on the choice of decomposition.
Below, we introduce two different decompositions and the respective relevant bipartite entropies [for a sketch, see Fig.~\ref{fig:intro}(a)].

\subsection{Operator entanglement}

First we consider a direct matrix-product form of $\hat \rho$ written as an MPDO as illustrated on the left-hand side of Fig.~\ref{fig:intro}(a)~\cite{zwolak2004mixed,orus2008infinite,werner2016positive,GuthJarkovsky_Effici_2020,weimer2021simulation}. For a concise description, we further combine the two physical indices on each site into a joint index to ``vectorize'' $\hat \rho$ and write
\begin{align}
\label{eq:mpdo_decomposition}
    \hat \rho = \sum_{\{i_n\}} \sum_{\{a_n\}}^\chi \prod_n R_{a_n a_{n+1}}^{[n]\,i_n} \lambda_{a_{n}}^{[n],\rho} \bigotimes_n \hat e_{i_n}.
\end{align}
The $R_{a_n a_{n+1}}^{[n]\,i_n}$ are three-dimensional tensors and the $\lambda_{a_{n}}^{[n],\rho}$ are vectors of normalized Schmidt values [$\sum_{a_n}(\lambda_{a_{n}}^{[n],\rho})^2=1$]. As  common in matrix-product decompositions, only the $\chi$ largest Schmidt values are retained to obtain an approximate representation of $\hat \rho$~\cite{zwolak2004mixed,orus2008infinite}. The basis operators $\hat e_{i_n}$ need to be chosen orthonormal, $\Tr(\hat e_i \hat e_j) = \delta_{ij}$. A commonly used choice is a straightforward linearization with $\hat e_{1} = \ket{0}\bra{0}$, $\hat e_{2} = \ket{0}\bra{1}$, $\hat e_{3} =\ket{1}\bra{0}$, and $\hat e_{4} =\ket{1}\bra{1}$. This choice has the disadvantage that the basis operators are not Hermitian, and as a consequence the entries of the $R$ tensors are generally complex. 
We point out that instead one can also keep the tensors real by choosing e.g.,~the Hermitian generalized Gell-Mann matrices~\cite{bertlmann2008bloch} as basis operators, which for local physical dimension 2 are related to the Pauli matrices as
\begin{align}
    \hat e_{1} = \frac{1}{\sqrt{2}}\id  \quad &
    \hat e_{2} = \frac{1}{\sqrt{2}}\hat \sigma^x\nonumber \\
    \hat e_{3} = \frac{1}{\sqrt{2}}\hat \sigma^y \quad &
    \hat e_{4} = \frac{1}{\sqrt{2}}\hat \sigma^z. \label{eq:ggm_basis}
\end{align}
Having to deal with only real numbers can be a practical advantage in terms of memory and run time. However, this basis may prevent implementations of algorithms that exploit conservation of magnetization in the XXZ model. The Pauli matrix basis has, e.g.,~also been employed in tensor network simulations in Ref. ~\cite{wojtowicz2020open}. Alternatively a density matrix-product operator matrix can also be defined in a locally purified form~\cite{werner2016positive}, which has the advantage of retaining the positivity of the density matrix after bond dimension truncation.

\medskip

Analogous to pure state MPS decompositions, the required bond dimension $\chi$ necessary to faithfully represent a state can be connected to the entropy in a bipartition of the chain~\cite{schuch2008entropy,GuthJarkovsky_Effici_2020}. In the case of density matrices, this leads to the definition of OE
\begin{align} \label{entropy1}
    S_{\rm OP}^{(n)} = - \sum_{a_{n}} \left(\lambda_{a_{n}}^{[n], \rho}\right)^2 \log_2\left(\lambda_{a_{n}}^{[n],\rho}\right)^2\, .
\end{align}
Since the OE can be generally largest in the center of the chain, we only consider $S_\mathrm{OP} \equiv S_\mathrm{OP}^{(N/2)}$. It is important to point out that for mixed states, $S_{\rm OP}$ is not necessarily a genuine measure of entanglement. However, it remains a crucial quantity linked to the efficiency of the decomposition~\eqref{eq:mpdo_decomposition}, since for a truncation to bond dimension $\chi$, the OE is limited to values ${S_{\rm OP}} \leq \log_2 (\chi)$.

For our model in the thermodynamic limit $N\rightarrow \infty$, the density matrix remains invariant with respect to translations by two lattice sites. Numerical simulations can exploit this symmetry to effectively simulate infinite chains using an an infinite time-evolving block decimation (iTEBD) algorithm with proper re-orthogonalization~\cite{orus2008infinite,wellnitz2022rise}. Except for Fig.~\ref{fig:intro}, in this paper we make use of this and all MPDO simulations will be for $N=\infty$.

\medskip

In Fig.~\ref{fig:intro}(b) we show an example evolution of the OE dynamics in a weak-dissipation limit ($N=40$, $\gamma_\pm=0.01J$, $\gamma_z=0$). For times at which Hamiltonian dynamics dominates ($tJ \ll 1/\gamma_\pm$) the OE initially exhibits a quick linear growth, such that the simulation quickly saturates to the maximum possible OE imposed by the finite bond dimension, $S_{\rm OP} \leq \log_2(\chi)$ (horizontal dashed lines). For example, for $\chi = 128$, a faithful simulation of the dynamics with the representation of Eq.~\eqref{eq:mpdo_decomposition} is only possible up to times of $tJ \sim 3$. For small dissipation, MPDO simulations thus become numerically inefficient quickly.

\subsection{Trajectory entanglement }
Alternatively, the density matrix $\hat \rho$ can be decomposed into a statistical mixture of pure states as sketched on the right hand side of Fig.~\ref{fig:intro}(a). For example, a quantum trajectory evolution algorithm (see Sec.~\ref{sec:method_qt}) produces an approximation of $\hat \rho$ of the form:
\begin{align}
    \hat \rho \approx \frac{1}{N_t} \sum_{\tau=1}^{N_t} \ket{\psi_\tau}\bra{\psi_\tau}.
    \label{eq:dm_statmix}
\end{align}
Each of the $N_t$ states $\ket{\psi_\tau}$ can then be decomposed as an ordinary MPS in canonical form~\cite{vidal2004efficient}:
\begin{align}
\label{eq:mps-qt}
    \ket{\psi_\tau} = \sum_{\{i_n\}} \sum_{\{a_n\}}^\chi \prod_n \Gamma_{a_n a_{n+1}}^{[n]\,i_n} \lambda_{a_{n}}^{[n], \tau} \bigotimes_n \ket{{i_n}}\,.
\end{align} 
As in Eq.~\eqref{eq:mpdo_decomposition}, the $\Gamma_{a_n a_{n+1}}^{[n]\,i_n}$ are three-dimensional tensors and the $\lambda^{[n],\tau}$ are real-valued Schmidt vectors, which are truncated at a maximum bond dimension $\chi$. $\ket{i_n =1,2} = \ket{\downarrow,\uparrow}_n$ describes the local physical basis of spin $n$.

\begin{figure*}
	\centering
    \includegraphics[width=\linewidth]{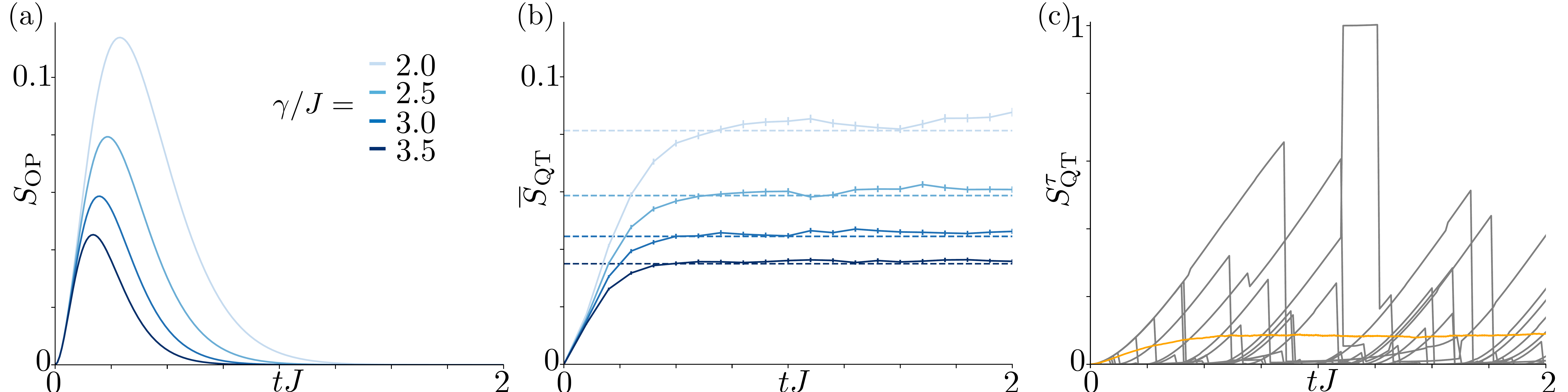}
	\caption{\textit{Strong spontaneous emission and absorption}--- (a/b) Time evolution of $S_{\text{OP}}$ (a) and $\overline{S}_{\text{QT}}$ (b) for various rates of $\gamma/J = 2,2.5,3, 3.5$ (blue solid lines, color from light to dark). $S_{\text{OP}}$ exhibits a clear rise and fall behavior, whereas $\overline{S}_{\text{QT}}$ quickly reaches a constant plateau value. Statistical error bars are barely visible in (b). Horizontal dashed lines in (b) correspond to the analytical estimate from Eq.~\eqref{eq:an_plateau}. (c) Dynamics of $S_{\rm QT}^\tau$ for several individual trajectories. Here, $\gamma/J = 2$, and $\overline{S}_{\text{QT}}$ is shown as a solid yellow line.  [For MPDO: $N=\infty$, $\chi = 512$; for QT+MPS: $N = 40$, $\chi = 32$, $N_t = 2000$].}
	\label{fig:decay_strong}
\end{figure*}

Analogous to pure state MPS, the amount of bipartite entanglement in each trajectory for a bond $n$ is quantified by von Neumann entropy
\begin{align}
    S_{\rm QT}^{\tau(n)} = -\sum_{a_{n}} \qty(\lambda_{a_{n}}^{[n], \tau})^2 \log_2 \qty(\lambda_{a_{n}}^{[n], \tau})^2,
    \label{eq:singletraj-entanglement}
\end{align}
and we will compute it around the center of the chain, averaging over 11 center sites to reduce the standard deviation (see Appendix Fig. \ref{fig:appendix3}). The MPS decomposition of Eq.~\eqref{eq:mps-qt} implies that $S^\tau_{\rm QT} \leq \log_2(\chi)$. We now define TE as the average entanglement over all trajectories
\begin{align} \label{entropy2}
    \overline{S}_{\rm QT} = \lim_{N_t \rightarrow \infty} \frac{1}{N_t} \sum^{N_t}_{\tau=1} S_{\rm QT}^\tau.
\end{align}
Importantly, just like OE, for mixed states TE is not necessarily a genuine measure of quantum entanglement, but it is still linked to the efficiency of the QT+MPS simulation. In addition to TE we will also analyze fluctuations of the TE, by computing TE sample standard deviations,
\begin{align}
    \sigma_{\overline S_{\text{QT}}} = \sqrt{\frac{1}{N_t-1} \sum_\tau \left( S^\tau_{\rm QT} - \overline{S}_{\rm QT}\right)^2}.
    \label{eq:stddev}
\end{align}
As error bars we use the sample standard error of the mean, $\epsilon = \sigma_{\overline S_{\text{QT}}}/\sqrt{N_t}$. Note that the individual trajectories are not translation invariant, such that we cannot use an infinite MPS method to work in the thermodynamic limit.

In Fig.~\ref{fig:intro}(c) we show an example evolution of the TE for the same setup as in in Fig.~\ref{fig:intro}(b), i.e.~in the weak-dissipation limit [$N=40$, $\gamma_\pm=0.01J$, $\gamma_z=0$]. Here we used $N_t=400$, and the entropy is averaged over 11 bonds in the center (see Appendix~\ref{sec:app_convergence} for a discussion on this bond averaging). Like the OE, the TE increases approximately linearly at short times, and then saturates to a plateau value (horizontal dashed line) imposed by $\chi$. However, the TE increases only half as fast as the OE in Fig.~\ref{fig:intro}(b) due to the local dimension only being half the size. Therefore, with QT+MPS we can faithfully simulate dynamics  up to twice the simulation time, $tJ \sim 6$. Thus, for small dissipation rates it seems evident that QT+MPS is preferable over MPDO. Not only is the local state space smaller, but also the bipartite entropy growth is reduced. However, below we will see that the situation changes drastically for larger dissipation rates.

\section{Comparing MPDO and QT+MPS}

We now systematically compare entropy growth in MPDO and QT+MPS simulations also for larger-noise cases. We start by pointing out some practical advantages of the MPDO approach. As discussed in previous work~\cite{bonnes2014superoperators}, quantum trajectory approaches are best suited for small dissipation rates, since the most common implementation is first order in the time step $\gamma \Delta t$, where only one jump can happen per time step per decay channel. Extensions to higher orders~\cite{steinbach1995high} are cumbersome to implement for many-body systems with many decay channels. As a consequence, for large dissipation rates, time steps get prohibitively small. In contrast, the MPDO approach relies on a Trotter expansion for which well established higher-order methods exist~\cite{Sornborger_Higher_1999} (e.g.~here we use a fourth-order method, see Appendix~\ref{sec:app_convergence}). In addition, individual trajectories can break symmetries that are present in the full density matrix, such as translation invariance in the case considered here. The latter can be used to significantly boost the efficiency of the MPDO simulations~\cite{orus2008infinite,wellnitz2022rise}. Finally, the stochastic nature of trajectories only allows us to resolve observables with some noise for a fixed number of trajectories $N_t$. Thus, to accurately determine the absolute magnitude of an observable $\overline A$ requires a large number of trajectories $N_t \gg \sigma_A^2/ \overline A^2$ ~\cite{press1992art}, where $\overline A$ and $\sigma_A^2$ are the average and variance of the observable $A$.

\medskip

Besides those practical considerations, the most important difference between both methods is the growth of the corresponding bipartite entropy, OE for MPDO and TE for QT+MPS, since the simulation complexity scales exponentially with entropy. In Sec.~\ref{sec:decay} and Sec.~\ref{sec:dephasing} we  analyze the entropy dynamics in the presence of spontaneous emission and absorption, and dephasing, respectively. For QT+MPS not only the averaged entanglement of trajectories, but also the variations of entanglement are important, which we discuss in Sec.~\ref{sec:stddev}.

\subsection{Spontaneous emission and absorption}
\label{sec:decay}

In this section we consider spontaneous emission and absorption, i.e.~dynamics with Lindblad operators of Eqs.~\eqref{eq:Lp} and \eqref{eq:Lm} for $\gamma_-=\gamma_+\equiv \gamma$. In Fig.~\ref{fig:decay_strong} we start by considering a large dissipation scenario with $\gamma > J$. 

\smallskip

In Fig.~\ref{fig:decay_strong}(a), we observe that the OE dynamics is initially (at times $t \ll \gamma^{-1}$) exhibiting a quick build-up, which can be attributed to dominating Hamiltonian dynamics in this short-time limit. Approximating the state at short times as a pure state, $\hat \rho(t) \approx \ket{\psi_t} \bra{\psi_t}$ with $\ket{\psi_t} = e^{-i\hat{H}t}\ket{\psi_0}$, it becomes obvious that the OE growth is twice as fast as the entanglement entropy growth in $\ket{\psi_t}$ (see, e.g., Ref. \cite{dubail2017entanglement}). The latter dominates the short-time dynamics of the TE in Fig.~\ref{fig:decay_strong}(b). At times $t \gtrsim \gamma^{-1}$, however, the initial coherence is quickly destroyed by the dissipation, and the OE decreases creating a well-known ``rise and fall'' behavior~\cite{carollo2021emergent,alba2021hydrodynamics}. The long-time value of $S_{\rm OP}=0$ can be understood, as the overall entropy increase with time pushes the system towards a trivial maximum entropy state $\hat \rho \propto \id$ with $S_\mathrm{OP} = 0$.

Fig.~\ref{fig:decay_strong}(b) shows the evolution of $\overline{S}_{\text{QT}}$ for the same parameters. Again, at times $t \ll \gamma^{-1}$ there is an initial rise, but at longer times, in contrast to $S_{\text{OP}}$, rather than going to zero, $\overline{S}_{\text{QT}}$ reaches a plateau with a value depending on $\gamma$. This implies that at sufficiently long times  $S_{\rm OP} < \overline{S}_{\text{QT}}$. This means that the density matrix is then decomposed as statistical mixture \eqref{eq:dm_statmix} of trajectories with (on average) finite entanglement. QT+MPS therefore produces an inefficient state representation of the trivial density matrix $\hat \rho \propto \id$  and therefore requires bond dimensions $\chi > 1$, in contrast to the MPDO approach.

\smallskip

To better understand the origin of the plateau value of $\overline{S}_{\text{QT}}$ in the large dissipation case, in Fig.~\ref{fig:decay_strong}(c) we show $S_{\rm QT}^\tau$ for a few individual trajectories. It is striking that in this regime, nearly every quantum jump leads to a reset of the entanglement to zero. More precisely, there are three dominating scenarios that can be identified in the evolution of $S_{\rm QT}^\tau$. After an initial rise, i) $S_{\rm QT}^\tau$ drops to zero followed by a rise, ii) $S_{\rm QT}^\tau$ drops to zero and remains at zero, and iii) $S_{\rm QT}^\tau$ jumps to 1. All of these scenarios can be understood analytically from a simple model.

\begin{figure}
	\centering
    \includegraphics[width=\linewidth]{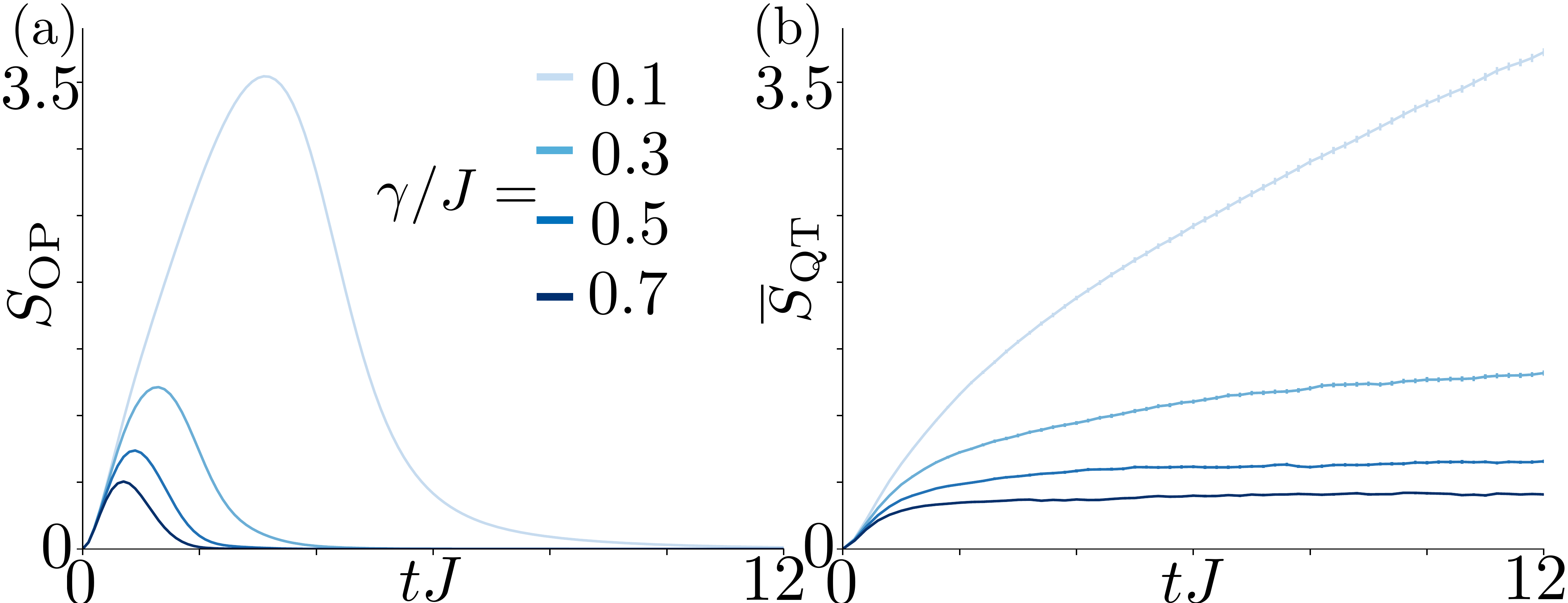}
	\caption{\textit{Intermediate spontaneous emission and absorption rates }--- (a/b) Time evolution of $S_{\text{OP}}$ (a) and $\overline{S}_{\text{QT}}$ (b) for various rates of $\gamma/J = 0.1,0.3,0.5, 0.7$ (blue solid lines, color from light to dark). $S_{\text{OP}}$ exhibits a clear rise and fall behavior, whereas, for the smallest values of $\gamma$, $\overline{S}_{\text{QT}}$ follows a continuous increase that decreases with the increase of $\gamma$ until reaching a plateau around values of $\gamma/J\geq 0.5$. Statistical error bars are barely visible in (b). [$\chi = 256$; for MPDO: $N=\infty$, for QT+MPS: $N = 40$, $N_t = 500$].}
	\label{fig:decay_low}
\end{figure}

Consider a system of just two neighboring spins initially in the state $\ket{\downarrow\uparrow}$. At sufficiently short times, Hamiltonian dynamics evolves the system into a state $\propto (\ket{\downarrow\uparrow} + \delta \ket{\uparrow\downarrow})$ with some small amplitude $|\delta| \ll 1$. This state features finite $S_{\rm QT}^\tau > 0$, which however will be immediately destroyed by a quantum jump of the form \eqref{eq:Lp} or \eqref{eq:Lm}. Post-jump states are then either $\ket{\downarrow\uparrow}$ or $\ket{\uparrow\downarrow}$ [rare cases with probability of order $\mathcal{O}(\delta^2)$], which both will start to build up entanglement again [scenario i)], or more likely [with probability of $\mathcal{O}(1)$] they will be $\ket{\downarrow\downarrow}$ or $\ket{\uparrow\uparrow}$, for which Hamiltonian~\eqref{eq:xxz} does not induce entanglement [scenario ii)] until the next quantum jump. Note that although for the two-spin case discussed here technically only scenario ii) is relevant, other relevant cases enter when more than two sites are involved. For example, to understand scenario iii), it is crucial to consider an effective block of four neighboring spins (see Sec.~\ref{sec:analytical}). Taking this into account allows us to derive an analytical result including terms up to order $\mathcal{O}(J^2/\gamma^2)$ for the plateau value, which is given in Eq.~\eqref{eq:an_plateau}. These analytical estimates are shown as horizontal dashed lines in Fig.~\ref{fig:decay_strong}(b), and fit the numerical result especially especially well for large $\gamma$. The fact that $S_{\rm QT}^\tau$ can also jump to $1$ emphasizes again the importance of also considering the strong fluctuations of TE in QT+MPS, which we will analyze in more detail in Sec.~\ref{sec:stddev}.

\medskip

For intermediate dissipation rates $\gamma \lesssim J$, the relevant physics cannot be described by simple few-spin arguments anymore. In this case we find that $\overline{S}_{\text{QT}}$ in QT+MPS simulations starts to exhibit a steady growth on the time-scale of multiple spin-exchange interactions $tJ \sim 10$ for $\gamma \lesssim 0.3 J$ [see Fig.~\ref{fig:decay_low}(b)]. Importantly, as expected, the evolution of $S_{\rm OP}$ still exhibits the usual rise and fall behavior [see Fig.~\ref{fig:decay_low}(b)]. Taking the example of $\gamma = 0.1J$ in Fig.~\ref{fig:decay_low}, this implies that $\overline{S}_{\text{QT}}$ overcomes the equivalent $S_{\rm OP}$ (after the fall) already for $tJ \sim 6$, and $\overline{S}_{\text{QT}}$  continues to grow. It eventually also surpasses the peak value of $S_{\rm OP}$.  Here, for $\gamma=0.1J$ we observe a continuous growth of $\overline{S}_{\text{QT}}$ on our whole simulation time-scale. While from our simulations there is no definite proof that this growth would continue for $\chi \to \infty$ and $N \to \infty$, the results are e.g.~in line with recent predictions for entanglement transitions in measured systems for similar fermionic hopping and spin models~\cite{alberton2021entanglement,azad2021phase}. However, in this context it is interesting to emphasize that also in a case where $\overline{S}_{\rm QT}$ keeps growing (e.g.~for $\gamma = 0.1$ and the range $4 \lesssim tJ \leq 12$), the OE is already decaying towards zero.

\medskip

\begin{figure}[bt]
	\centering
    \includegraphics[width=\linewidth]{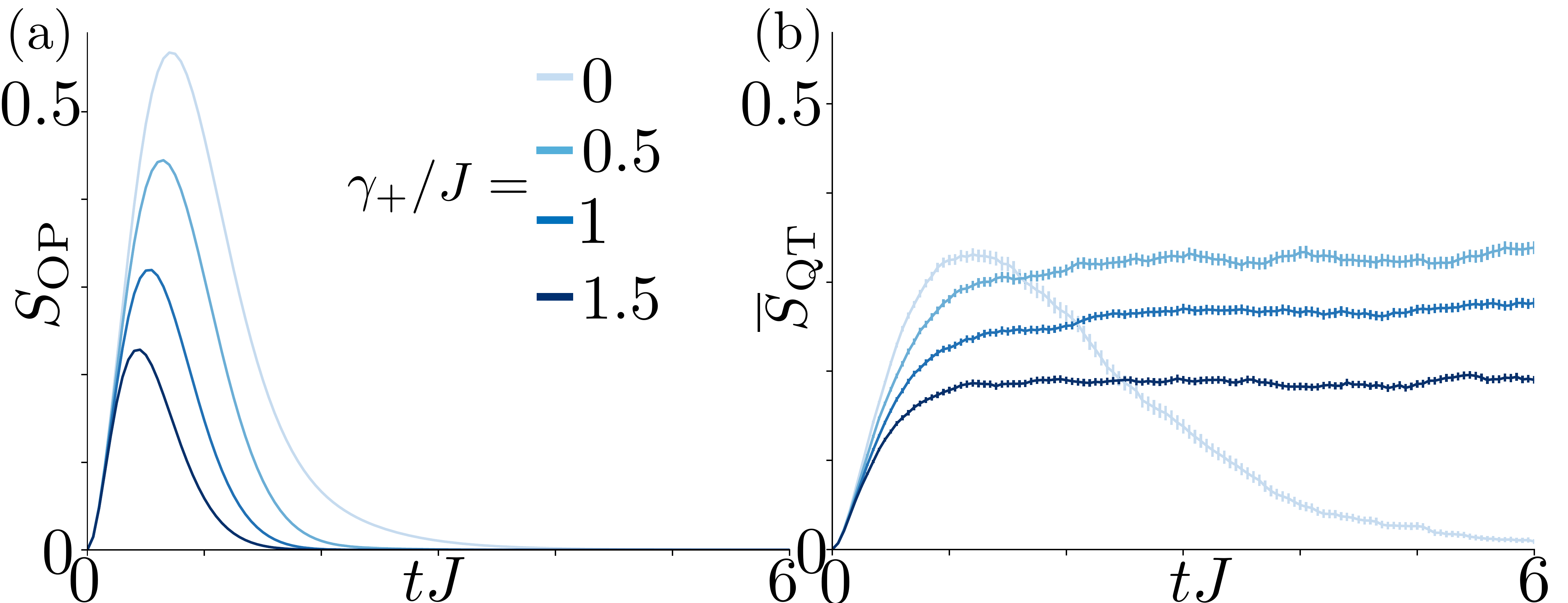}
	\caption{\textit{Imbalanced $\gamma_-\neq \gamma_+$}--- Time evolution of entropies with fixed $\gamma_- = J$ and $\gamma_+/J = 0,0.5,1,1.5$ (blue solid lines, color from light to dark). (a) For $S_{\rm OP}$ the rise and fall behavior persists and the peak OE only decreases with $\gamma_+$. (b) For $\overline{S}_{\text{QT}}$ the plateau height generally decreases with $\gamma_+$. Only for $\gamma_+/J = 0$, $\overline{S}_{\text{QT}} \to 0$.  [For MPDO: $N=\infty$, $\chi=256$; for QT+MPS: $\chi = 64, N_t  = 800$].}
	\label{fig:decay_imbalanced}
\end{figure}

Finally, in Fig.~\ref{fig:decay_imbalanced} we also show the effect of the experimentally more relevant scenario of imbalanced emission and absorption rates, $\gamma_- \neq \gamma_+$. In MPDO simulations we generally find that dynamics exhibits a robust rise and fall behavior with the maximum OE simply depending on the overall amount of dissipation, $\gamma_-+\gamma_+$ [Fig.~\ref{fig:decay_imbalanced}(a)]. Similarly, the plateau value found for $\overline{S}_{\text{QT}}$ decreases with increasing $\gamma_+$ when $\gamma_-=J$ fixed [Fig.~\ref{fig:decay_imbalanced}(b)]. Only for values very close to $\gamma_+ \approx 0$, the quantum jumps start to remove almost all excitations from the system. Therefore, $\overline{S}_{\text{QT}}$ only vanishes in the long time limit for $\gamma_+=0$, when the system evolves into the trivial product steady state $\bigotimes_i \ket{\downarrow}_i$.

\subsection{Dephasing}
\label{sec:dephasing}

\begin{figure*}[tb]
	\centering
    \includegraphics[width=\linewidth]{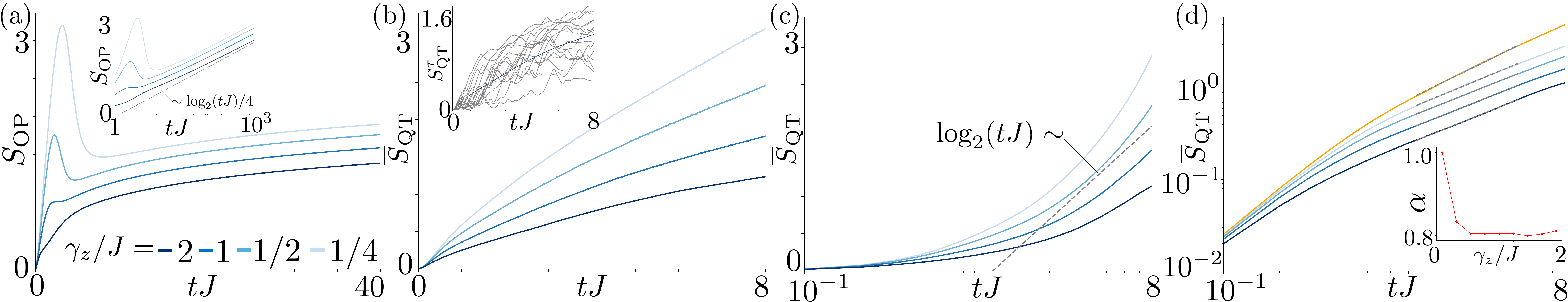}
	\caption{\textit{Dephasing}--- Time evolution of entropies for $\gamma_z/J = 0.25, 0.5, 1, 2$ (blue solid lines, color from light to dark). (a) Rise and fall of the OE, and universal logarithmic long-time growth, $S_{\rm OP} = \frac{1}{4}\text{log}_2(tJ) + \text{const.}$ (inset: log-lin scale). Reproducing results in Ref.~\cite{wellnitz2022rise}.
 (b) In contrast, $\overline{S}_{\text{QT}}$  growth steadily in QT+MPS simulations. The inset shows ${S}_{\text{QT}}^\tau$ for several trajectories and $\gamma_z=2J$. (c) ${S}_{\text{QT}}^\tau$ on a lin-log scale, demonstrating super-logarithmic growth in contrast to $S_{\rm OP}$ in panel (a). (d) $\overline{S}_{\text{QT}}$ on a log-log scale. Solid yellow line: $\gamma_z = 0$. Thin dashed lines show power law fits $\overline{S}_{\text{QT}} \propto t^{\alpha}$. Fitting range is chosen to best fit $\alpha=0$ (linear increase) for $\gamma_z=0$, $1.1 < tJ < 3.9$. Inset: Fitted power-law exponent as function of $\gamma_z$. [For MPDO: $N=\infty$, $\chi=512$; for QT+MPS: $N=40$, $\chi = 256$, $N_t = 400$].}
	\label{fig:dephasing}
\end{figure*}

In this section we consider dephasing as noise model, i.e.~dynamics with Lindblad operators of the form~\eqref{eq:Lz}. Fig.~\ref{fig:dephasing} summarizes our findings.

\smallskip

Fig.~\ref{fig:dephasing}(a) reproduces the evolution of $S_{\rm OP}$ from Ref.~\cite{wellnitz2022rise}. It is shown that after the rise and fall the OE exhibits a slow universal logarithmic increase as $\log_2(tJ)/4$ (see inset for the long-time evolution on a lin-log scale). In Ref.~\cite{wellnitz2022rise} it is shown that this growth can be traced back to a classical stochastic sub-diffusion process of symmetry blocks in the density matrix, and is inherent to the master equation and the initial state being $U(1)$ symmetric.

In contrast, in Fig.~\ref{fig:dephasing}(b) we now also analyze the corresponding evolution of $\overline{S}_{\text{QT}}$ in a QT+MPS simulation for identical parameters as in panel (a). It is striking that no rise and fall behavior is present; instead $\overline{S}_{\text{QT}}$ grows continuously while the absolute values are decreased with increasing dephasing rate $\gamma_z$. 

The inset in Fig.~\ref{fig:dephasing}(b) shows a small number of individual trajectories. In contrast to Fig.~\ref{fig:decay_strong}(c), here quantum jumps do not lead to discontinuous dynamics, i.e.~they do not reset ${S}_{\text{QT}}^\tau$ to zero. This can be again rationalized from a simple two-spin argument in the short-time limit. Considering the short-time approximation to the state, being $\propto (\ket{\downarrow\uparrow} + \delta \ket{\uparrow\downarrow})$, it is clear that a unitary quantum jump of the form~\eqref{eq:Lz} leads to a state with flipped relative phase $\propto (\ket{\downarrow\uparrow} \pm \delta \ket{\uparrow\downarrow})$, but cannot quench entanglement. Here, entanglement growth must be explained by a more complex interplay between Hamiltonian and Lindbladian dynamics, reflected in the continuous variation of individual trajectories in the inset.

\smallskip

Comparing OE evolution in Fig.~\ref{fig:dephasing}(a) with TE evolution in Fig.~\ref{fig:dephasing}(b), an important fact to point out is that TE growth is significantly faster for time-scales $t \gtrsim 1/\gamma_z$. In Fig.~\ref{fig:dephasing}(c) we plot $\overline S_{\text{QT}}$ on a log-lin scale, to clearly demonstrate a super-logarithmic increase. This implies that for $tJ \gtrsim 1/\gamma_z$, TE quickly becomes larger than OE in MPDO simulations.

\smallskip

In Fig.~\ref{fig:dephasing}(d) we plot  $\overline S_{\text{QT}}$ on a log-log scale and find that the growth of TE at intermediate time scales (of a few tunneling processes $J^{-1}$) resembles a power law $\overline S_{\text{QT}} \propto ct^{\alpha}$ with $\alpha$ barely depending on $\gamma_z$. To obtain an estimate of $\alpha$ we perform a power-law fit in the window $1.1 < tJ < 3.9$ (dashed lines). This time window is chosen to avoid on the one hand short-time physics, and on the other hand finite-size effects at long times. As criterion for a good fitting range we chose the range where the fit reproduces the expected exponent $\alpha = 1$ (linear increase) with the smallest possible error in the case of $\gamma_z=0$. The $\gamma_z=0$ simulation is shown as solid yellow line in Fig.~\ref{fig:dephasing}(d). The inset in Fig.~\ref{fig:dephasing}(d) shows the fitted values for $\alpha$ as function of $\gamma_z$. Remarkably, for $\gamma_z \gtrsim 0.5J$ we observe a quite robust value of $\alpha \approx 0.8$. It has to be emphasized, however, that this finding is restricted to a transient time-range due to the relatively small system size ($N=40$) accessible in our simulation. We therefore do not draw conclusions about whether this effect persists universally over extended time scales.
It would be an interesting prospect to investigate this closer in the future in larger systems and for longer times.

\subsection{Trajectory entanglement fluctuations in QT+MPS}
\label{sec:stddev}

\begin{figure}[b]
	\centering
    \includegraphics[width=\columnwidth]{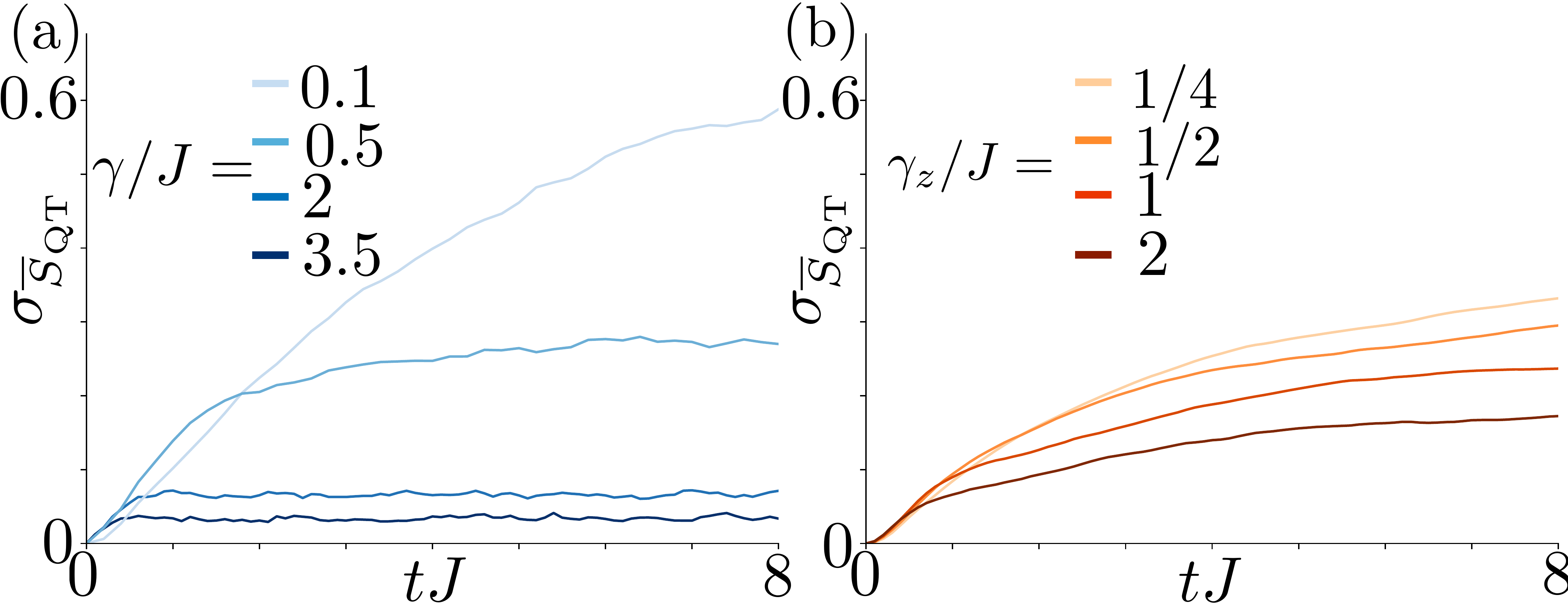}
	\caption{\textit{Standard deviation of TE} --- Time evolution of sample standard deviations of the TE. (a) With spontaneous emission and absorption for both strong and weak dissipation rates $\gamma/J = 3.5, 2.0, 0.5, 0.1$ (blue solid lines, color from light to dark) [$\chi = 32$ and $N_t = 2000$]. (b) With dephasing for various rates $\gamma_z/J = 0.25, 0.5, 1, 2$ (orange solid lines, color from light to dark) [$\chi = 256$ and $N_t = 400$]. In both cases, the growth of $\overline{S}_{\text{QT}}$ is correlated with a growth of its standard deviation $\sigma_{\overline{S}_{\text{QT}}}$ over time [in comparison with~Figs.~\ref{fig:decay_strong}(b), \ref{fig:decay_low}(a), \ref{fig:dephasing}(b)].}
	\label{fig:stddev}
\end{figure}

In this section we finally also show results on the variations of the entanglement entropy in individual trajectories of our QT+MPS simulations. In Fig.~\ref{fig:stddev}(a) and Fig.~\ref{fig:stddev}(b) we plot the sample standard deviation, $\sigma_{\overline S_{\text{QT}}}$, for the data of our previous simulations with spontaneous emission and absorption, and for dephasing, respectively. There, an important insight is that the growth of $\overline S_{\text{QT}}$ is also correlated with the growth of its standard deviation $\sigma_{\overline S_{\text{QT}}}$. In cases where $\overline{S}_{\rm QT}$ reaches a constant value quickly, e.g.~for $\gamma \gtrsim 2.0$ [see Fig. \ref{fig:decay_strong}(b)], also $\sigma_{\overline S_{\text{QT}}}$ saturates quickly in Fig.~\ref{fig:stddev}(a). In contrast, in all other cases where  $\overline S_{\text{QT}}$ shows a persistent growth on our simulation time-scales, i.e.~for $\gamma/J \lesssim 0.3$ [see Fig. \ref{fig:decay_low}(b)] and for all considered values of $0.25 \leq \gamma_z/J \leq 2$ [see Fig. \ref{fig:dephasing}(b)], also the standard deviation $\sigma_{\overline S_{\text{QT}}}$ increases over time. Since the bond dimensions need to be chosen large enough to include ``worst-case'' trajectories with entropies $\sim S^\tau_{\rm QT} + \sigma_{\overline S_{\text{QT}}}$, the presence of an increasing standard deviation implies that bond dimensions should be much larger than expected from the averaged entropy $\overline S_\mathrm{QT}$ only.

\section{Details on numerical and analytical calculations}
\label{sec:methods}

In this section we provide and discuss some technical details of our MPDO and QT+MPS algorithms and provide the analytical calculation for the plateau values observed in Fig.~\ref{fig:decay_strong}.

\subsection{MPDO Simulations}

Equations of motion for our MPDO decomposition in~\eqref{eq:mpdo_decomposition} can be easily derived and depend on the choice of the basis matrices $\hat e_{i_n}$.
Inserting a general many-body density matrix expansion in the chosen basis,
\begin{align}
    \hat{\rho} 
    = 
    \sum_{i_1,i_2,\dots,i_N} r_{i_1,i_2,\dots,i_N} \hat e_{i_1} \hat e_{i_2} \dots \hat e_{i_N},
\end{align}
into the master equation~\eqref{eq:master}, leads to an evolution equation which only depends on nearest-neighbor terms for the vectorized density matrix elements, i.e.
\begin{align}
\frac{d}{dt} r_{i_1,i_2, \dots, i_N} = \sum_{k_n, k_{n+1}}\left( A_{i_n,i_{n+1}}^{k_n, k_{n+1}}\right) r_{i_1,\dots,k_n,k_{n+1}\dots, i_N},
\end{align}
for sites $n$ and $n+1$.

The relevant super-operator two-site gates $A_{i_n, i_{n+1}}^{k_n, k_{n+1}}$ include Hamiltonian and Lindblad terms. Explicitly, in the bulk of the chain,
\begin{align}
 A_{i_n, i_{n+1}}^{k_n, k_{n+1}} =  H_{i_n, i_{n+1}}^{k_n, k_{n+1}} 
 + \frac{1}{2} \left( L_{i_n}^{k_n} \delta_{i_{n+1}}^{k_{n+1}} + \delta_{i_{n}}^{k_{n}}  L_{i_{n+1}}^{k_{n+1}} \right),
 \label{eq:supop_gates}
\end{align}
with $\delta_{i_{n}}^{k_{n}}$ being Kronecker deltas. The Hamiltonian and Lindblad super-operator expansions $H_{i_n, i_{n+1}}^{k_{n},k_{n+1}}$ and $L_{i_{n}}^{k_{n}}$ can be easily computed numerically: From the XXZ Hamiltonian~\eqref{eq:xxz}, 
$
\hat H_{
\rm XXZ} \equiv \sum^{N-1}_{n=1} \hat H_{i_n, i_{n+1}}
$
via
\begin{align}
 H_{i_n, i_{n+1}}^{k_{n},k_{n+1}} = 
 \tr\left[ \hat e_{i_n} \hat e_{i_{n+1}} 
 \left(-\mi [\hat H_{i_n, i_{n+1}}, \hat e_{k_{n}} \hat e_{k_{n+1}}] \right)
  \right],
\end{align}
and for the Lindblad terms via
\begin{align}
    L_{i_{n}}^{k_{n}} 
    &= 2\tr{\hat e_{i_n} \hat L_\eta  \hat e_{k_n} \hat L_\eta^\dag }\nonumber \\
    &\quad - \tr{\hat e_{i_n} \hat L_\eta \hat e_{k_n}}
    - \tr{\hat e_{i_n} \hat e_{k_n} \hat L_\eta }
 \end{align}
with the corresponding jump operators $\hat L_\eta$ from Eqs.~\eqref{eq:Lm},~\eqref{eq:Lp}, or~\eqref{eq:Lz}.

\smallskip

We compute the two-site superoperator gates~\eqref{eq:supop_gates} and apply them to the MPDO decomposition~\eqref{eq:mpdo_decomposition} using a standard TEBD~\cite{vidal2004efficient} algorithm. As common, we can use conservation laws to significantly speed up gate applications by exploiting the block-diagonal form of the super-operator gates $A_{i_n, i_{n+1}}^{k_n, k_{n+1}}$. For dephasing this is enabled by the fact that the gates commute with the expansion of the super-operator for the total magnetization, $\hat S_z \hat \rho$. It is interesting to point out that even for spontaneous emission and absorption the gates commute with a superoperator of the form $[\hat \rho, \hat S_z]$, which can be exploited. 

\smallskip

We also exploit translational invariance of our model to simulate dynamics using an iTEBD algorithm, in which case due to the non-Hermitian nature of the super-operator gates we resort to the re-orthogonalization scheme presented in Ref. ~\cite{orus2008infinite}. We also emphasize again that numerical truncation errors in the TEBD step lead to a loss of positivity of the density matrix. This can in principle be avoided by using a locally purified representation of the density matrix as proposed in Ref. ~\cite{werner2016positive}. In practice however, since we always operate in a regime of negligible truncation (results are converged in $\chi$, see Appendix~\ref{sec:app_convergence}), we do not find the loss of positivity to be a numerical problem.

\subsection{Quantum Trajectories}
\label{sec:method_qt}

For our QT+MPS simulations we use a variation of a standard  QT method~\cite{dalibard1992wave,dum1992monte,carmichael2009open}, also known as the quantum Monte Carlo wave-function method, quantum jump approach, or simulated number measurement unraveling~\cite{daley2014quantum,weimer2021simulation,vovk2021entanglement}. In its original formulation a trajectory $\ket{\psi_\tau(\Delta t)}$ is evolved with the non-Hermitian effective Hamiltonian, $\hat H_{\rm eff}$, over a time $\Delta t$, and a Lindblad jump operator is randomly applied with a probability proportional to $1-\|\ket{\psi_\tau(\Delta t)}\|^2$, followed by a renormalization. The specific Lindblad operator $\hat L_{\eta_0}$ for  a decay channel $\eta_0$ is chosen with probability $\propto \bra{\psi_\tau(t)} \hat L_{\eta_0}^\dag \hat L_{\eta_0} \ket{\psi_\tau(t)}$.

In this original formulation, the method suffers from the inherent problem that the equivalence of the master equation relies on a first-order Euler argument~\cite{daley2014quantum} and thus requires small time-step sizes $\Delta t$. This problem can be partly ameliorated by an alternative formulation of QT, in which quantum jump times are numerically pre-calculated from the decay-dynamics of the norm, such that the system can be evolved to the quantum jump time with an arbitrarily high-order method~\cite{daley2014quantum, dum1992monte}.

\smallskip

We make use of this idea, but in our case also exploit the fact that the non-Hermitian part of our effective Hamiltonian has a particularly simple form:
\begin{align}
        \hat H_{\rm nh} = -\frac{\mi}{2} \sum_{\eta} \hat L^{\dag}_\eta \hat L_\eta =-\frac{\mi N\gamma }{2}\mathbb{1}
\end{align}
for both spontaneous emission and absorption with equal rate, $\gamma = \gamma_\pm$, and  dephasing with $\gamma= \gamma_z$. Trivially, this also implies $[\hat H_{\text{XXZ}}, \hat H_{\text{nh}}] = 0$. The time-dependent decay of the squared norm thus follows
\begin{align}
    \|\ket{\psi_\tau(t)}\|^2 &=\bra{\psi_\tau(t=0)} e^{-2i \hat H_{\text{nh}} t}\ket{\psi_\tau (t=0)}, \nonumber\\
    &=e^{-N \gamma t}. \label{eq:decaynorm}
\end{align}
This implies that the time for a jump $j+1$,  $t_{j+1}$, can be easily pre-calculated~\cite{wolff2020numerical}, e.g.~for a random threshold value $0 \leq r_{j+1} \leq 1$ with $ t_{j+1} = t_j - \text{ln}(r_{j+1})/N\gamma$.

It is important to point out, however, that also this approach fundamentally suffers from the issue that the difference between two consecutive jumps is $(t_{j+1}-t_{j}) \propto 1/N\gamma$. Thus the selected time step for the time evolution is on average limited by $\Delta  t \sim 1/N\gamma$ and has to decrease with system size $N$. This strongly limits the applicability of the method for large $N$.

For large $N$, a way around this problem can be found in cases where jump operators commute, $[\hat L_\eta, \hat L_\mu]=0$ for $\eta \neq \mu$. Then, multiple quantum jumps can be applied within a single time step. Therefore one can first evolve $\ket{\psi_\tau}$ with only the Hermitian Hamiltonian $\hat H_{\rm XXZ}$, and then conditionally apply quantum jumps after non-Hermitian Hamiltonian dynamics for individual decay channels $\eta$, ${-\mi \hat L_{\eta}^\dag \hat L_{\eta}/2}$, as e.g.~described in Ref. ~\cite{vovk2021entanglement}. While this scheme still leads to a first-order method, the method then only introduces time-step errors of order $\order{(\Delta t \gamma)^2}$, instead of $\order{(\Delta t N\gamma)^2}$. For higher-order integration schemes one should resort to schemes as presented in Ref. ~\cite{steinbach1995high}, which are however cumbersome to implement in case of a large amount of decay channels.

\subsection{Analytical Formula for a plateau}
\label{sec:analytical}

In this section, we develop a further understanding of  the plateau value of $\overline{S}_{\rm QT}$ observed in quantum trajectory simulations with spontaneous emission and absorption at sufficiently long times. Our reasoning is based on simple analytical arguments using blocks of just a few spins.

\smallskip

We start by considering only two spins, $N=2$. In a many-body system, these two spins may represent the spins adjacent to the bipartite splitting in the center of the chain. In a regime of sufficiently large dissipation (such that entanglement cannot build up over distances larger than two spins), the overall entanglement will thus be dominated by the entanglement entropy of one of the two single spins. For two spins, there are four distinct possible initial states $ \{ \ket{\uparrow \uparrow}, \ket{\downarrow \downarrow},  \ket{\uparrow \downarrow}, \ket{\downarrow \uparrow } \}$. The states $ \{ \ket{\uparrow \uparrow}, \ket{\downarrow \downarrow} \}$ are left unchanged by the dynamics under Hamiltonian \eqref{eq:xxz}, which implies that the entanglement entropy remains strictly zero in this case. In contrast, the initial ``N\'eel states'', $\ket{\phi_1} \equiv \ket{\uparrow \downarrow}$  and $\ket{\phi_2} \equiv \ket{\downarrow \uparrow}$ evolve under Hamiltonian~\eqref{eq:xxz}. Starting. e.g., with ~$\ket{\phi_1}$ the state at time $t$ is 
\begin{align}
\ket{\psi(t)} 
= \frac{1}{\sqrt{2}} \left( \me^{-\mi E_- t} \ket{E_-} + \me^{-\mi E_+ t} \ket{E_+} \right),
\label{eq:2spin_evo}
\end{align}
with $E_\pm  =  -J/4 \pm J/2$ two eigenvalues of \eqref{eq:xxz} for $N=2$ and the corresponding two eigenstates are
$\ket{E_\pm} = \left(\ket{\phi_1} \mp \ket{\phi_2} \right)/\sqrt{2}.$
Then, the reduced density matrix for the single spin of state \eqref{eq:2spin_evo} evolves according to
\begin{align}
&\rho_1(t) = \cos^{2}(tJ/2)\ket{\uparrow}\bra{\uparrow} + \sin^{2}(tJ/2)\ket{\downarrow}\bra{\downarrow},
\end{align}
which leads to the single-spin entanglement entropy evolution
\begin{align}
S_1(t) =&-\cos^{2}(tJ/2) \log_2(\cos^{2}(tJ/2)) \nonumber \\
	&- \sin^{2}(tJ/2) \log_2(\sin^{2}(tJ/2)).
	\label{eq:2_sites_entropy}
\end{align}

A quantum jump event can happen after a certain time $t$ with a probability given by Eq.~\eqref{eq:decaynorm}. Hence, the probability distribution function for a jump occurring on any of the two sites is given by an exponential distribution of the form
\begin{equation}
	f_2(t) = 2\gamma \me^{- 2t\gamma} \label{eq:dist},
\end{equation}
with $\gamma = \gamma_+ = \gamma_-$, and a mean time between jumps of $\bar \tau = {1}/{2\gamma}$. Importantly, from the jump operators \eqref{dis}, we readily see that for $N=2$ a quantum jump will systematically collapse $\ket{\psi(t)}$ into one of the initial product states $ \{ \ket{\uparrow \uparrow}, \ket{\downarrow \downarrow},  \ket{\uparrow \downarrow}, \ket{\downarrow \uparrow } \}$, resetting the entanglement entropy to zero. As argued already in Sec.~\ref{sec:decay}, starting in a state $\ket{\phi_1}$ or $\ket{\phi_2}$, for jumps at a short time of $\sim 1/2\gamma$, the collapsed state will almost always be $\ket{\downarrow\downarrow}$ or $\ket{\uparrow\uparrow}$. Vice versa, on average after a time of $1/2\gamma$ the states $\ket{\downarrow\downarrow}$ and $\ket{\uparrow\uparrow}$ will collapse into $\ket{\phi_1}$ or $\ket{\phi_2}$. Thus on average we can assume that in a steady-state situation and large $\gamma$, all four states are equally probable. Therefore, using a short-time expansion of \eqref{eq:2_sites_entropy} and keeping terms of order $\mathcal{O}(J^2/\gamma^2)$, we can derive
\begin{align}
	\overline{S}^{2-\text{site}}_{\rm QT} &= \frac{1}{2} \int_0^{\infty} S(t) f_J(t) \, \mathrm{d}t  \label{eq:an_avent}\\
    &\approx  \frac{1}{2} \frac{\gamma J^{2} }{ 2 \ln(2)} \int_0^{\infty}  t^{2} e^{-2\gamma t}\big[1 - \text{ln}((Jt/2)^2)\big] \mathrm{d}t\\ 
    &=  \frac{ J^{2} }{16\gamma^{2} \ln(2)} \big[ 2(E - 1) + \ln((\gamma^2 16)/J^2) \big]
	\label{eq:an_aventsol},
\end{align}
where $E$ is the Euler-Mascheroni constant. The $1/2$ in \eqref{eq:an_avent} stems from the fact that only half of the states contribute to the entropy growth.

While this result gets close to the long-time TE in Fig.~\ref{fig:decay_strong}(b), we did not find a perfect match with the numerical solution in the large $\gamma$ limit. We attribute this to the fact that there are additional contributions of order $\mathcal{O}(J^2/\gamma^2)$ coming from rare jump events in a four-spin block that collapse to states with TE $1$, as seen, e.g., ~in Fig.~\ref{fig:decay_strong}(c). To illustrate how such a scenario can arise, consider an initial state of the form $\ket{\downarrow}\ket{\phi_1}\ket{\uparrow}$. At short times $t \ll 1/J$ the state is approximated by
\begin{align}
\ket{\phi'} \approx \ket{\downarrow \uparrow \downarrow \uparrow} - \mi\frac{t J}{2} (\ket{\uparrow \downarrow \downarrow \uparrow} + \ket{\downarrow \downarrow \uparrow \uparrow} + \ket{\downarrow \uparrow \uparrow \downarrow}).
\label{eq:pre_rare_state}
\end{align}
Denoting the probability for a jump on spin $i$ by $p_{i,\pm} \propto \bra{\phi'} \hat \sigma_i^\mp \hat \sigma_i^\pm \ket{\phi'}$, the probability for a jump with $\hat \sigma^+_2$ (on the second spin) is $p_{2,+} \propto J^2/\gamma^2$. However, the state after this jump will be given by:
\begin{align}
\hat \sigma^+_2\ket{\phi'} \propto (\ket{\uparrow \uparrow \downarrow \uparrow} + \ket{\downarrow \uparrow \uparrow \uparrow}).
\label{eq:post_rare_state}
\end{align}
This is an entangled state between blocks of sites $(1,2)$ and $(3,4)$ with entropy $1$. Since such a state will remain robust on a $1/\gamma$ time-scale, overall such a process will give another contribution to the plateau value of the TE of order $\mathcal{O}(J^2/\gamma^2)$. Note that the identical argument holds for a jump with $\hat \sigma_3^-$, and for jumps with  $\hat \sigma_2^-$ and $\hat \sigma_3^+$ if we would consider the initial N\'eel state $\ket{\uparrow}\ket{\phi_2}\ket{\downarrow}$. 

\smallskip

Using this, we can now roughly estimate the correction due to such processes in steady state. We consider again jumps on the two central spins and focus on the initial state $\ket{\downarrow}\ket{\phi_1}\ket{\uparrow}$. Up to quadratic order in $t$, $p_{2,+}(t) \approx t^2 J^2/4$ [after using state~\eqref{eq:pre_rare_state} and normalizing with four possible jump probabilities $p_{2,3,\pm}$]. 
This then leads to a correction term of 
\begin{align}
    \overline{S}_{\rm QT}^{\rm corr} = \frac{1}{4} \int_0^\infty \frac{t^2 J^2}{4} f_2(t) \, \mathrm{d}t = \frac{J^2}{32\gamma^2} \label{eq:an_4spincorr}
\end{align}
Here, the factor $1/4$ comes from the fact that we assume that in steady state a N\'eel state configuration in the four-site block has the probability of $1/8$, and the fact that we have two possible ``entangling jumps'' per N\'eel state. Summing the contributions from Eq.~\eqref{eq:an_aventsol} and \eqref{eq:an_4spincorr} we arrive at our final estimate for the plateau value of
\begin{align}
\overline{S}_{\text{QT}} &=\frac{ J^{2} }{16\gamma^{2} \ln(2)} \big[ 2(E - 1) + \ln((\gamma^2 16)/J^2)\big] + \frac{J^2}{32 \gamma^2}.
 \label{eq:an_plateau}
\end{align}
This estimate provides an excellent estimation for the numerical results in the large $\gamma$ limit as shown in Fig.~\ref{fig:decay_strong}(b).

\section{Conclusion and Outlook}

\label{sec:concl}

Here, we made a systematic comparison between two approaches to numerical simulations of open quantum many-body dynamics with matrix-product decompositions. We denote them (i) MPDO (evolving full matrix-product density operators) and (ii) QT+MPS (``number measurement unraveling'' of the density matrix into MPS trajectories). In our paper we focused on a comparison of bipartite entropy growth dynamics in the two methods: operator entanglement for the MPDO approach, and trajectory entanglement for QT+MPS. For both methods these entropies are fundamentally linked to the respective numerical efficiency of the state representation. We compared the entropy growth behavior for two types of single-spin dissipative mechanisms: spontaneous emission and absorption, and dephasing.  

\smallskip

On a technical level, we discussed several  details of both numerical approaches and highlight several advantages of the MPDO method compared to QT+MPS. It allows for (i) Easier implementation of high-order Trotter decompositions (thus larger time steps), (ii) easier exploitation of translational invariance, and (iii) a real-valued representation of the tensors in the matrix-product decomposition.

On a fundamental level, our most important insight is that for time-scales $\gtrsim 1/\gamma$ (with dissipative rates $\gamma$), the  growth of OE is generally slower than that of TE. For spontaneous emission and absorption OE generally vanishes at long times. In contrast, we find that TE grows on our simulable time-scales for small $\gamma$, and approaches a constant value quickly for large $\gamma$ (for which we found an analytical estimate). For dephasing, OE grows only logarithmically~\cite{wellnitz2022rise}, while TE (and also its sample standard deviation) grows as a power law. For our (number measurement) unraveling, we found a robust growth exponent of $\sim 0.8$ for sufficiently large dephasing rates. In both cases our results imply that at sufficiently long times, MPDO matrices become a numerically significantly more efficient state representation compared to the ones obtained from QT+MPS. In other words, the standard QT approach produces a statistical mixture of entangled trajectories that can classically be less inefficiently represented than the whole density matrix.

\smallskip

Our simulations make use of a simple initial product state. In the future it would be interesting to also include more complex entangled initial states for systematic studies, which have, e.g., ~led to a clearly decreased efficiency of the MPDO approach in Ref. ~\cite{wolff2020numerical}. 

Furthermore, another interesting prospect would be to study in depth the variations of TE growth under different types of environmental measurement schemes (e.g.~homodyne measurement unravelings~\cite{vovk2021entanglement}). The dynamics of trajectory resolved entanglement in continuously monitored open systems has been a very recent topic of study in the context of ``measurement-induced phase transitions''~\cite{skinner2019measurement,cao2019entanglement,fuji2020measurement,alberton2021entanglement,turkeshi2021measurement,azad2021phase}. It is an interesting future prospect to investigate whether the scaling behavior that we observed here (e.g.~the robust power-law behavior for dephasing) is a feature of the most commonly used number measurement unraveling, or if it also holds for different unraveling schemes.

\medskip

\section*{Acknowledgments} We thank Vincenzo Alba and J\'er\^ome Dubail for very helpful input on operator entanglement, as well as Guido Pupillo, Shannon Whitlock, Tomaž Prosen, Marko Žnidaric and Enej Ilievski for valuable discussions. This work was supported by LabEx NIE Contract No. ANR-11-LABX0058 NIE, and the QUSTEC program, which has received funding from the European Union's Horizon 2020 research and innovation program under the Marie Sk\l{}odowska-Curie Grant No. 847471. This work is part of the Interdisciplinary Thematic Institute QMat, as part of the ITI 2021-2028 program of the University of Strasbourg, Centre National de la Recherche Scientifique and Inserm, and was supported by IdEx Unistra (Grant No. ANR-10-IDEX-0002), SFRI STRAT'US project (Grant No. ANR-20-SFRI-0012), and EUR QMAT Grant No. ANR  -17-EURE-0024 under the framework of the French Investments for the Future Program. D.W.~acknowledges funding by NIST. Our codes make use of the intelligent tensor library (ITensor)~\cite{itensor}. Computations  were  carried  out  using  resources  of  the High Performance Computing Center of the University of Strasbourg, funded by Equip@Meso (as part of the Investments for the Future Program) and CPER Alsacalcul/Big Data.

\appendix

\begin{widetext}
  
\section{Details on numerical convergence}
\label{sec:app_convergence}

In this appendix we present details on the numerical convergence for the results presented in the main text. We discuss convergence in the matrix-product bond dimension $\chi$, convergence in the time-step sizes, and our bond-averaging procedure that we use in QT+MPS. Throughout the main text we only show results that are converged in both $\chi$ and times-step sizes.

\subsection{Bond dimension convergence}

In order to make sure that our simulations are converged in the bond dimension $\chi$ for both representations [MPDO and QT+MPS defined in  Eqs.~\eqref{eq:mpdo_decomposition} and \eqref{eq:mps-qt}, respectively], we repeat our simulations while doubling $\chi$ for each run until lines become visually indistinguishable. All parameters used for the figures in the main text are summarized in Table~\ref{tab:numparams} below. For spontaneous emission and absorption we varied the bond dimension in the range $\chi = 32, 64, 128, 256$ and for dephasing in the range $\chi = 64, 128, 256, 512$. The convergence plots in Fig.~\ref{fig:appendix1} demonstrate this procedure for the scenario where OE and TE grow largest (i.e.~for cases where the largest $\chi$ values are necessary). We choose the parameters from Fig.~\ref{fig:decay_low} (a) and (b) for MPDO and QT+MPS with $\gamma = 0.1$ simulated up to times $tJ = 12$. In the case of MPDO, the OE reaches a peak of value $S_{\text{OP}} \approx 3.5$, while in the case of QT+MPS, as explained in section \ref{sec:decay}, the dissipation $\gamma$ shows a continuous increase reaching a value of $\overline{S}_{\text{QT}} \approx 3.5$ around times $tJ = 12$. Note that for $\chi =256$, fundamentally an entropy of $\log_2(256)=8$ could be supported, and in practice we observe convergence for $\chi \gtrsim 128$ in both cases.  For dephasing, to reach convergence of $\chi$ we choose parameters from Fig.~\ref{fig:dephasing} (a) and (b), for MPDO and QT+MPS respectively, with $\gamma_z/J = 0.25$. In the case of MPDO the OE reaches its highest value in the short-time peak $S_{\text{OP}} \approx 3.0$, and convergence is clearly reached for $\chi\gtrsim 256$. For QT+MPS, just as in the case of weak spontaneous emission and absorption the TE shows a continuous growth reaching a value of $\overline{S}_{\text{QT}} \approx 3.0$ around $tJ = 8$, and convergence is already reached for $\chi\gtrsim 64$.

\begin{figure}[t]
	\centering
    \includegraphics[width=\linewidth]{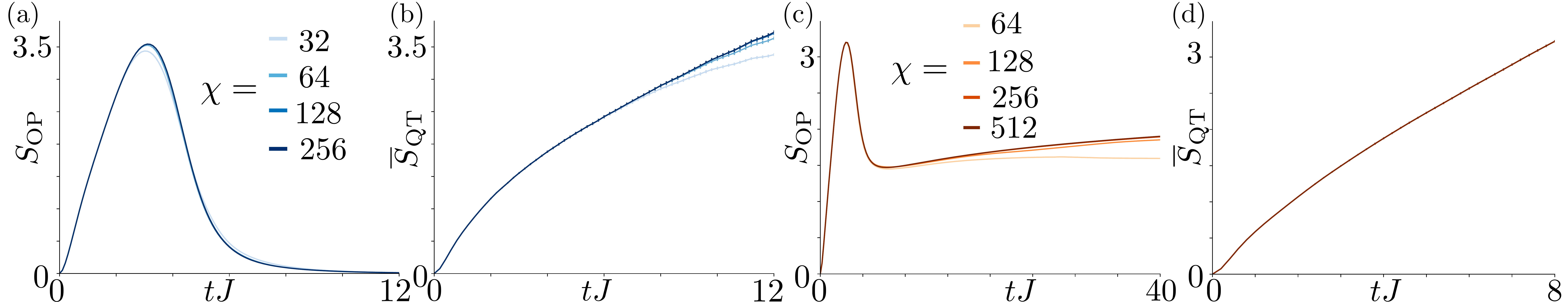}
	\caption{\textit{Convergence in $\chi$} --- For (a/b) we consider spontaneous emission and absorption with a rate $\gamma/J = 0.1$ for different values of $\chi = 32, 64, 128, 256$. We notice that a value of $\chi = 128$ suffices for convergence at all times for $S_{\text{OP}}$ and for $\overline{S}_{\text{QT}}$ it suffices until $tJ \approx 12$ . For (c/d) we consider dephasing with a rate $\gamma_z/J = 0.25$ for different values of $\chi = 64, 128, 256, 512$. In this case a value of $\chi = 256$ suffices at least until $tJ = 40$ for the case of $S_{\text{OP}}$. For the case of $\overline S_{\text{QT}}$, at least until $tJ = 8$, a value of $\chi = 64$ is enough to reach convergence [for MPDO: $N=\infty$, for QT+MPS: $N=40$, $N_t = 500$ (a/b), $N_t = 400$ (c/d)].}
	\label{fig:appendix1}
\end{figure}

\begin{figure}[t]
	\centering
    \includegraphics[width=\linewidth]{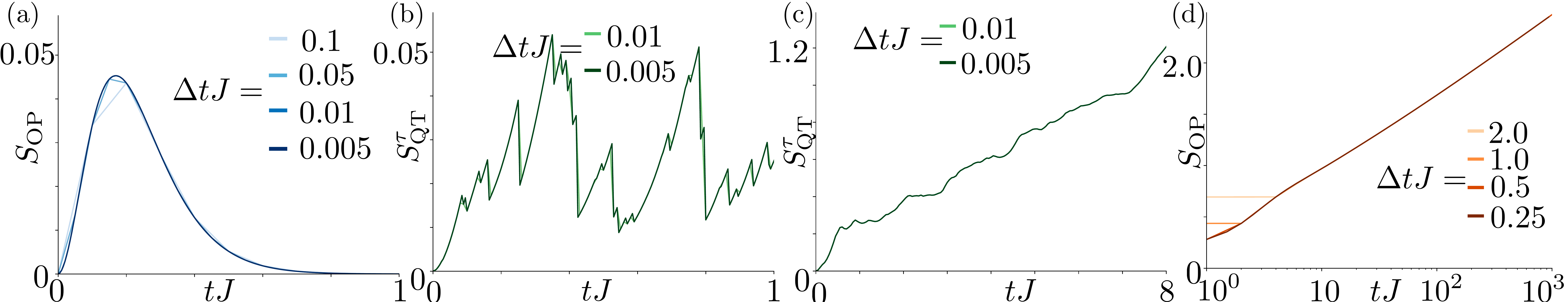}
	\caption{\textit{Time step convergence} --- For (a/b) we consider spontaneous emission and absorption with a large rate $\gamma/J = 3.5$. In the case of $S_{\text{OP}}$ a value of $\Delta t = 0.1/J$ already clearly suffices. In our QT+MPS methods Trotter time steps ($h$) are adaptive with $h \leq \Delta t$ to resolve jumps. However results also agree when $\Delta t$ becomes smaller than the inter-jump times (see text). For (c/d) we consider dephasing with a rate $\gamma_z/J =$2. In the case of $S_{\text{OP}}$ a time-step $\Delta t J = 2$ is already clearly enough due to a fourth order Trotter decomposition used in our code [$\chi = 512$; for MPDO $N=\infty$; for QT+MPS $N=40$].}
	\label{fig:appendix2}
\end{figure}

\begin{table}[h]
\begin{center}
    \begin{tabular}{c|c|c|c|c|c|c}
    
        Figure & $\gamma_+/J$ &$\gamma_-/J$ & $\gamma_z/J$ &$\chi$ & $\Delta t J$ & $N_t $\\
        \hline
        Fig.~1(b) &$0.01$& $0.01$ &0 & 8,16,32,64,128 & $0.1$ & ---\\
        Fig.~1(c) &$0.01$& $0.01$ &0 & 8,16,32,64,128 & $0.01$ & 400 \\
        Fig.~2(a) &$2$, $2.5$, $3$, $3.5$& $\gamma_+/J$&0 & 512 & $0.01$ & --- \\
        Fig.~2(b) &$2$, $2.5$, $3$, $3.5$& $\gamma_+/J$&0 & 32 & $0.01$ & 2000 \\
        Fig.~2(c) &$2$& 2 &0 & 32 & $0.01$ & 10 \\
        Fig.~3(a) &$0.1$, $0.3$, $0.5$, $0.7$& $\gamma_+/J$&0 & 256 & $0.1$ & --- \\
        Fig.~3(b) &$0.1$, $0.3$, $0.5$, $0.7$& $\gamma_+/J$&0 & 256 & $0.2$ & 500 \\
        Fig.~4(a) &0, $0.5$, 1, $1.5$& 1&0 & 256 & $0.05$ & --- \\
        Fig.~4(b) &0, $0.5$, 1, $1.5$& 1&0 & 64 & $0.005$ & 800 \\
        Fig.~5(a) &0& 0&$0.25$, $0.5$, 1,2 & 512 & $0.1$ & --- \\
        Fig.~5(a) (inset)&0& 0&$0.25$, $0.5$, 1,2 & 1024 & $0.1$, $1.0$ & --- \\
        Fig.~5(b) &0& 0&$0.25$, $0.5$, 1,2 & 256 & $0.1$ & 400 \\
        Fig.~5(c) &0& 0&$0.25$, $0.5$, 1,2 & 256 & $0.1$ & 400 \\
        Fig.~5(d) &0& 0&$0.25$, $0.5$, 1,2 & 256 & $0.1$ & 400 \\
        Fig.~5(d) (inset) &0& 0&$[0.25:0.25:2]$& 256 & $0.1$ & 400 \\
        Fig.~6(a) &$0.1$, $0.5$, $2$, $3.5$& $\gamma_+/J$&0 & 32 & $0.01$ & 2000 \\
        Fig.~6(b) &0& 0&$0.25$, $0.5$, 1,2 & 256 & $0.1$ & 400 \\
        Fig.~S1(a) & $0.1$ & $0.1$ & 0 & 32, 64, 128, 256 & 0.1 & --- \\
        Fig.~S1(b) &$0.1$ & $0.1$ & 0 & 32, 64, 128, 256 & 0.2 & 500\\
        Fig.~S1(c) & 0 & 0 & $0.25$ & 64, 128, 256, 512 & $0.2$ & ---\\
        Fig.~S1(d) & 0 & 0 & $0.25$ & 64, 128, 256, 512 & $0.2$ & 400 \\
        Fig.~S2(a) & $3.5$ & $3.5$ & 0 & 256 & $0.1, 0.05, 0.01, 0.005$ & ---\\
        Fig.~S2(b) & $3.5$ & $3.5$ & 0 & 256 & $0.1, 0.05, 0.01, 0.005$ & 1\\
        Fig.~S2(c) & 0 & 0 & 2 & 256 & $0.1, 0.05, 0.01, 0.005$ & ---\\
        Fig.~S2(d) & 0 & 0 & 2 & 256 &$0.1, 0.05, 0.01, 0.005$ & 1\\
        Fig.~S3(a)/(c) & 0 & 0 & 2.0 & 256 & 0.1 & 400\\
        Fig.~S3(b)/(d) & 3.5 & $3.5$ & 0 & 64 & 0.1 & 2000\\
       
    \end{tabular}
\end{center}
\caption{The numerical values of $\chi$, $\Delta t$, and $N_t$ used for all figures in this paper. Whenever two $\Delta t$ values are given, different $\Delta t$s were used for different parameters for historic reasons, but  are clearly converged in either case.}
\label{tab:numparams}
\end{table}

\subsection{Time-step convergence}

To check convergence in time-step size $\Delta t$, we also run simulations repeatedly, decreasing $\Delta t$ until lines become visually indistinguishable. To demonstrate this, in Fig.~\ref{fig:appendix2} we consider again ``worst-case'' scenarios with simulations for the largest rates $\gamma$ and $\gamma_z$ used in the main text (which generally require the smallest $\Delta t$). For all simulations in Fig.~\ref{fig:appendix2} we use $\chi=512$ such that the results are clearly converged in $\chi$. For the case of spontaneous emission and absorption  we use parameters from Fig \ref{fig:decay_strong} (a) and (b) with $\gamma = 3.5J$ for MPDO and QT+MPS, respectively. Here, for QT+MPS we show results for a single trajectory, but averaged over the 11 center bonds such that rapid jumps are still visible. For dephasing we use parameters from Fig.~\ref{fig:dephasing} (a) and (b) with $\gamma_z= 2 J$. In our QT+MPS algorithm we pre-calculate quantum jump times analytically and therefore the true time-step used in the Trotter decomposition gates, $h$ is adaptive and therefore $h \leq \Delta t$. For spontaneous emission/absorption (MPDO and QT+MPS) as well as for dephasing in QT+MPS [i.e.~in all panels Fig.~\ref{fig:appendix2}(a-c)] we vary $\Delta t$ in the range $\Delta t J=0.1, 0.05, 0.01, 0.005$, and only for dephasing and MPDO simulations [Fig.~\ref{fig:appendix2}(d)] we also demonstrate convergence for very large steps in the range $\Delta t J =2.0, 1.0, 0.5, 0.25$. The very large values of $\Delta t$ in MPDO are enabled by the fact that here we can easily incorporate a fourth-order Trotter decomposition for the matrix exponential of the master-equation super-operator. In particular, we use the following decomposition from Ref. \cite{Sornborger_Higher_1999}:
\begin{align}
    (1)^T(1)(1)^T(-2)(1)^T(1)^T(1)^T(1)^T(1)(1)^T(1)(1)(1)(1)(-2)^T(1)(1)^T(1),
\end{align}
where the notation ``$(n)^{(T)}$'' denotes a (transposed) sweep of two-spin super-operator gates with time-step $n\Delta t/12$. This allows us to observe convergence even for $\Delta t = 2J$ in Fig. ~\ref{fig:appendix2}(d). Also in Fig.~\ref{fig:appendix2}(a) the time step size $\Delta t \approx 0.1$ is already clearly sufficient. For QT+MPS in Fig.~\ref{fig:appendix2}(b) and (c) the mean times between jumps are on the order of $1/N\gamma \sim 0.007/J$ and $\sim 0.01/J$, respectively. Because of our adaptive time-step algorithm, the Trotter time step $h$ is on the order of those small values, and therefore a second-order decomposition is sufficient. Here we use the decomposition $(1)(1)^T$, where $(1)^{(T)}$~\cite{Sornborger_Higher_1999} denotes a (transposed) sweep with a time step of $h/2$ in the non-Hermitian two site gates. To show that this is sufficient we demonstrate that results are indistinguishable for $\Delta tJ = 0.01$ and $ 0.005$, where the latter is smaller than the mean inter-jump times in both cases. All time-step sizes used for the figures in the main text are also summarized in Table~\ref{tab:numparams} below.

\begin{figure}[t]
	\centering
    \includegraphics[width=\linewidth]{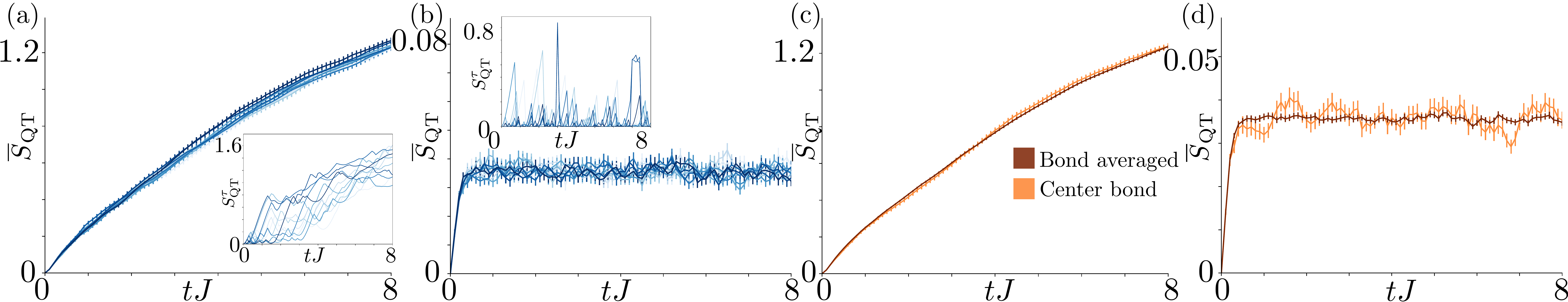}
    \caption{\textit{Reduction of standard error of the mean through bond averaging} --- (a/b) Trajectory averaged entropies  $\overline S_{\text{QT}}$ for the 11 central bonds (bond $16$ to $26$ from light to dark) for dephasing (a) and for spontaneous emission and absorption (b). The insets also show the entropy for the 11 bonds for a single trajectory. (c/d) Comparing the trajectory averaged entropy  $\overline S_{\text{QT}}$ for the center bond with the one averaged over the 11 center bonds. [for dephasing: $\gamma_z/J = 2.0$,  $\chi = 256$, and $N_t = 400$; for spontaneous emission and absorption: $\gamma/J = 3.5$, $\chi = 64$, and $N_t = 2000$].}
	\label{fig:appendix3}
\end{figure}

\begin{figure}[t]
	\centering
    \includegraphics[width=\columnwidth]{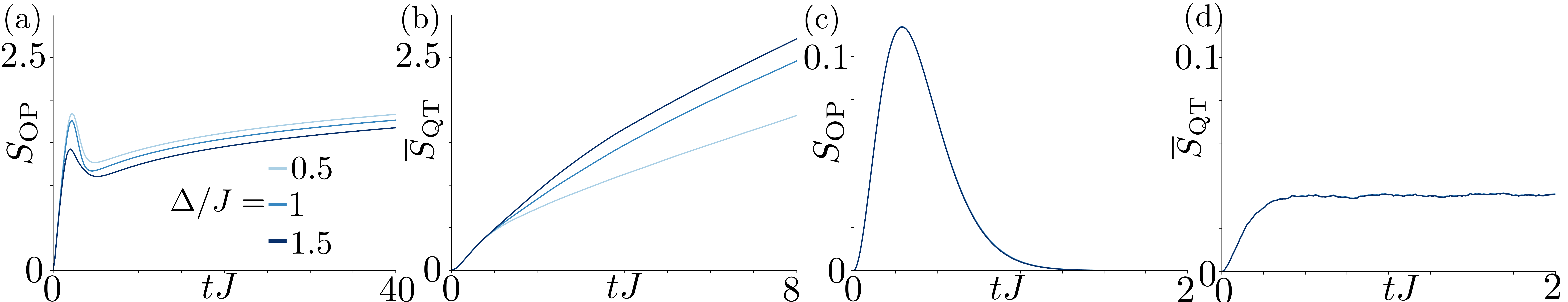}
\caption{\textit{Effects of ZZ spin-spin interaction $\Delta$} --- For (a/b) we consider dephasing with a rate $\gamma_z/J = 0.5$. For (c/d) we consider spontaneous emission and absorption with a rate $\gamma/J = 3.5$. In both cases we consider different values of $\Delta/J = 0.5, 1, 1.5$. [for MPDO: $N = \infty$, for dephasing we use $\chi = 1024$, for spontaneous emission and absoption we use $\chi = 512$; for QT+MPS: $N = 40$, for dephasing $N_t = 500$, $\chi = 1024$, for spontaneous emission and absortion $N_t = 2000$, $\chi = 256$]}
\label{fig:deltadif}
\end{figure}

\subsection{Bond averaging}

In Fig. \ref{fig:appendix3}, we verify that we can use bond averaging to reduce the sample standard error of the mean in the TE of QT+MPS simulations. In our system with $N=40$ we generally average the entropies over the five bonds neighboring the center bond, i.e., denoting the center bond as bond $21$, we average over all bonds from $16$ to $26$. In Fig.~\ref{fig:appendix3}(a) and (b) we show the trajectory averaged entropy $\overline{S}_{\rm QT}$ for dephasing (a) and spontaneous emission and absorption (b) for each of these bonds. Within statistical fluctuations given by the error bars (sample standard error of the mean) all entropies show similar behavior (the insets show variations for a single trajectory $S_{\rm QT}^\tau$, naturally exhibiting large fluctuations). This implies that the bond averaged entropy also agrees with the one of the center site, only with reduced error bars despite using the same number of trajectories. This is demonstrated in Fig.~\ref{fig:appendix3}(c) and (d) for dephasing and spontaneous emission and absorption, respectively.

\section{Effects of ZZ spin-spin interaction}
\label{sec:zz_dependence}

Here we consider an XXZ Hamiltonian, but with variable ZZ spin-spin interactions, $\Delta$, in the Hamiltonian:
\begin{align} \label{eq:xxzdelta}
\hat H_{
\rm XXZ} = \sum^{N-1}_{i=1}  \bigg[-\frac{J}{4}(\hat \sigma^x_i \hat \sigma^x_{i+1} + \hat \sigma^y_{i} \hat \sigma^y_{i+1}) + \frac{\Delta}{4} \hat \sigma^z_i \hat \sigma^z_{i+1}\bigg].
\end{align} 
We confirmed that a change of the sign $\Delta = +J \leftrightarrow \Delta = -J$ does not have any effect on the TE or OE evolution. Simulation results for different magnitudes $|\Delta|$ are shown in Fig.~\ref{fig:deltadif}. We point out the interesting fact that with increasing $\Delta$ the OE and the TE depend on $\Delta$ in opposite ways. While the OE decreases with increasing interaction strength [Fig.~\ref{fig:deltadif}(a)] (as also pointed out in the supplementary material of Ref. ~\cite{wellnitz2022rise}), the TE increases with $\Delta$. For the cases of large spontaneous and emission absorption, in contrast, the TE and OE remains essentially unchanged [Fig.~\ref{fig:deltadif}(c/d)].

\end{widetext}


%

\end{document}